# Primordial Nucleosynthesis in the Precision Cosmology Era

Gary Steigman

Departments of Physics and Astronomy, The Ohio State University, Columbus, Ohio 43210; email: steigman.1@osu.edu

## Abstract

Primordial nucleosynthesis probes the Universe during its early evolution. Given the progress in exploring the constituents, structure, and recent evolution of the Universe, it is timely to review the status of Big Bang Nucleosynthesis (BBN) and confront its predictions, and the constraints that emerge from them, with those derived from independent observations of the Universe at much later epochs. Following an overview of the key physics that controls the synthesis of the elements in the early Universe, the predictions of BBN in the standard (and some nonstandard) models of cosmology and particle physics are presented. The observational data used to infer the primordial abundances are described, with an emphasis on the distinction between *precision* and *accuracy*. These are compared with the predictions, testing the internal consistency of BBN and enabling a comparison of the BBN-inferred constraints with those derived from the cosmic microwave background radiation and large scale structure data.

**Contents**

# 1. INTRODUCTION

As the Universe evolved from its early, hot, dense beginnings (the Big Bang) to its present, cold, dilute state, it passed through a brief epoch when the temperature (average thermal energy) and density of its nucleon component were such that nuclear reactions were effective in building complex nuclei. Because the nucleon content of the Universe is small (in a sense to be described below) and because the Universe evolved through this epoch very rapidly, only the lightest nuclides (D, $^3$He, $^4$He, and $^7$Li) could be synthesized in astrophysically interesting abundances. The abundances of these relic nuclides provide probes of the conditions and contents of the Universe at a very early epoch in its evolution (the first few minutes) that would otherwise be hidden from our view. The standard model of cosmology subsumes the standard model of particle physics (specifically, three families of very light left-handed neutrinos, and their right-handed antineutrinos) and uses general relativity (e.g., the Friedman equation) to track the time evolution of the universal expansion rate and its matter and radiation contents. Big Bang Nucleosynthesis (BBN) begins in earnest when the Universe is a few minutes old and ends less than half an hour later when the nuclear reactions are quenched by the low temperatures and densities. The BBN-predicted abundances depend on the conditions at those times (e.g., temperature, nucleon density, expansion rate, neutrino content, neutrino-antineutrino asymmetry, etc.) and are largely independent of the detailed processes that established them. Consequently, BBN can test the standard models of cosmology and particle physics and constrain their parameters, as well as probe nonstandard physics or cosmology that may alter the conditions at BBN.

This review begins with a synopsis of the physics relevant to a description of the early evolution of the Universe at the epoch of primordial nucleosynthesis, and with an outline of the specific nuclear and weak-interaction physics of importance for BBN—in the standard model as well as in the context of some very general extensions of the standard models of cosmology and/or particle physics. Having established the framework, we then present our predictions of the relic abundances of the nuclides synthesized during BBN. Next, the current status of the observationally inferred relic abundances is reviewed, with an emphasis on the uncertainties associated with these determinations. The predicted and observed abundances are then compared to test the internal consistency of the standard model and to constrain extensions beyond the standard model. Observations of the cosmic microwave background radiation (CMB) temperature fluctuations and of the large scale structure (LSS) provide probes of the physics associated with the later evolution of the Universe, complementary to that provided by BBN. The parameter estimates/constraints from BBN and the CMB are compared, again testing the consistency of the standard model and probing or constraining some classes of nonstandard models.

## 2. A SYNOPSIS OF BIG BANG NUCLEOSYNTHESIS

The Universe is expanding and filled with radiation. All wavelengths, those of the CMB photons as well as the de Broglie wavelengths of all freely expanding massive particles, are stretched along with this expansion. Therefore, although the present Universe is cool and dilute, during its early evolution the Universe was hot and dense. The combination of high temperature (average energy) and density ensures that collision rates are very high during early epochs, guaranteeing that all particles, with the possible exception of those that have only gravitational strength interactions, were in kinetic and thermal equilibrium at sufficiently early times. As the Universe expands and cools, interaction rates decline and, depending on the strength of their interactions, different particles depart from equilibrium at different epochs. For the standard, active neutrinos ($\nu_e$, $\nu_\mu$, $\nu_\tau$), this departure from equilibrium occurs when the Universe is only a few tenths of a second old and the temperature of the photons, $e^\pm$ pairs, and neutrinos (the only relativistic, standard model particles populated at that time) is a few MeV. Note that departure from equilibrium is not sharp and collisions among neutrinos and other particles continue to occur. When the temperature drops below $T \lesssim 2$–3 MeV, the interaction rates of neutrinos with CMB photons and $e^\pm$ pairs become slower than the universal expansion rate (as measured by the Hubble parameter $H$), and the neutrinos effectively decouple from the CMB photons and $e^\pm$ pairs. However, the electron neutrinos (and antineutrinos) continue to interact with the baryons (nucleons) via the charged-current weak interactions until the Universe is a few seconds old and the temperature has dropped below ~0.8 MeV. This decoupling, too, is not abrupt; the neutrinos do not "freeze out" (see **Figure 1**).

Indeed, two-body reactions among neutrons, protons, $e^\pm$, and $\nu_e$ ($\bar{\nu}_e$) continue to influence the ratio of neutrons to protons, albeit not rapidly enough to allow the neutron-to-proton (n/p) ratio to track its equilibrium value of n/p $= \exp(-\Delta m/T)$, where the neutron-proton mass difference is $\Delta m = m_\mathrm{n} - m_\mathrm{p} = 1.29$ MeV. As a result,

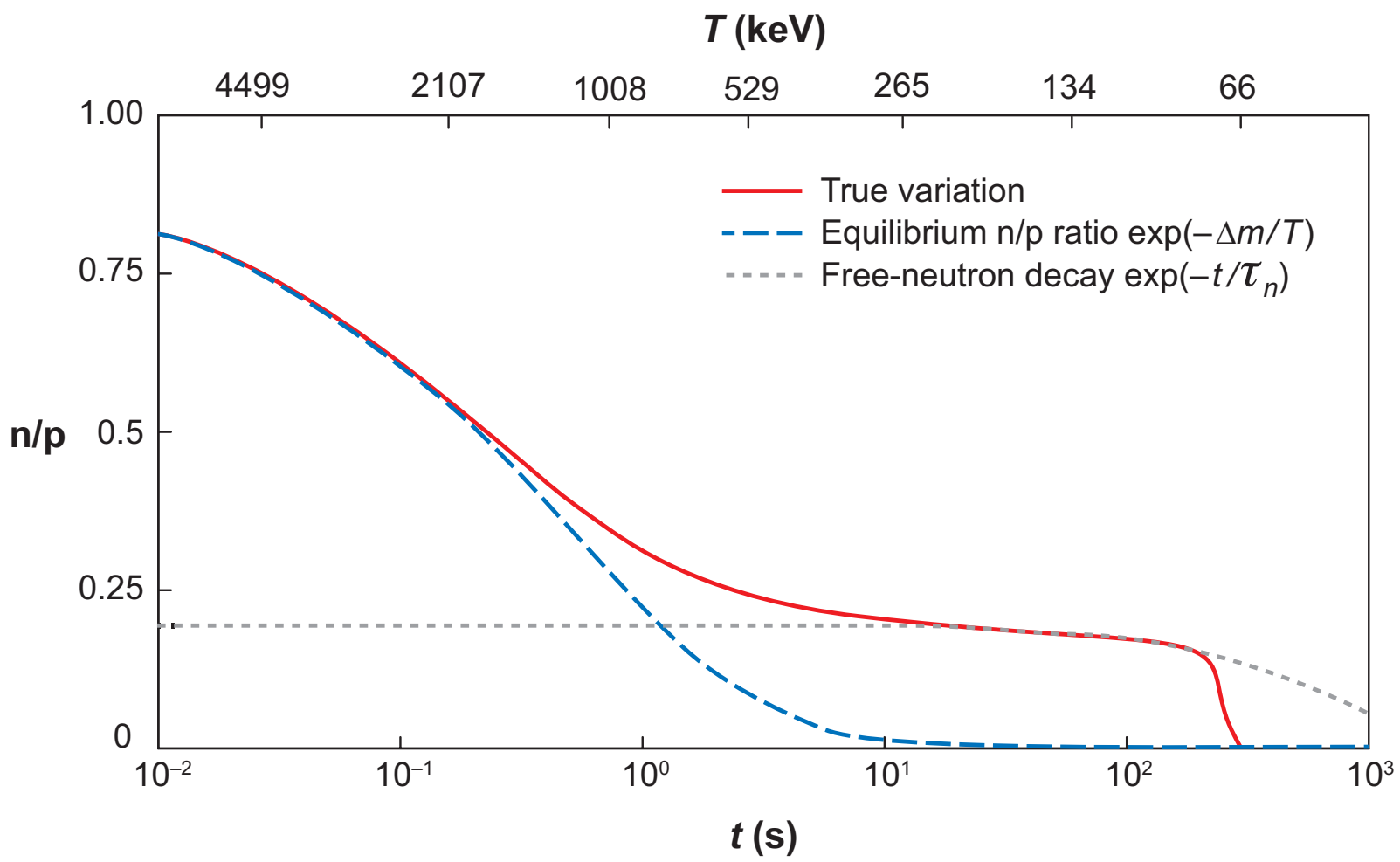

**Figure 1**

The time-temperature evolution of the neutron-to-proton (n/p) ratio. The solid red curve indicates the true variation. The steep decline at a few hundred seconds is the result of the onset of BBN. The dashed blue curve indicates the equilibrium n/p ratio $\exp(-\Delta m/T)$, and the dotted gray curve indicates free-neutron decay $\exp(-t/\tau_n)$.

the n/p ratio continues to decrease from ~1/6 at freeze out to ~1/7 when BBN begins at ~200 s ($T \approx 80$ keV). Because the neutrinos are extremely relativistic during these epochs, they may influence BBN in several ways. The universal expansion rate (the Hubble parameter $H$) is determined through the Friedman equation by the total energy density, which, during these early epochs, is dominated by massless particles in addition to those massive particles that are extremely relativistic at these epochs: CMB photons, $e^{\pm}$ pairs, and neutrinos. The early Universe is radiation dominated and the neutrinos constitute a significant component of the radiation. In addition, through their charged-current weak interactions, the electron-type neutrinos help control the fraction of free neutrons available, which, as we see below, effectively limits the primordial abundance of $^4$He.

Although the $e^{\pm}$ pairs annihilate during the first few seconds when $T \lesssim m_e$, the surviving electrons (the excess of electrons over positrons), equal in number to the protons to ensure charge neutrality, remain coupled to the CMB photons via Compton scattering. Only after the electrons and the nuclides (mainly protons and alphas) combine to form neutral atoms at "recombination" are the CMB photons released from the grasp of the electrons to propagate freely. This occurs when the Universe is some 400,000 years old, and the relic photons—redshifted to the currently observed black body radiation at $T = 2.725$ K—provide a snapshot of the Universe at this early epoch. At this relatively late stage (in contrast to BBN) in the early evolution of the Universe, the freely propagating, relativistic neutrinos contribute to the radiation density, influencing the magnitude of the universal expansion rate (e.g., the time-temperature relation). Note that if the neutrino masses are sufficiently large, the neutrinos will have become nonrelativistic prior to recombination, and their free streaming would have the potential to damp density fluctuations in the baryon fluid. The important topic of neutrino masses is not addressed here (for recent reviews see Reference 1); for our analysis, it is sufficient to assume that $m_\nu \lesssim$ few eV.

The primordial abundances of the relic nuclei produced during BBN depend on the baryon (nucleon) density and on the early-Universe expansion rate. The amplitude and angular distribution of the CMB temperature fluctuations depend on these same parameters (as well as on several others). The universal abundance of baryons may be quantified by comparing the number density of baryons (nucleons) to the number density of CMB photons,

$$\eta_{10} \equiv 10^{10}(n_{\rm B}/n_{\gamma}). \qquad 1.$$

As the Universe expands, the densities of baryons and photons both decrease, whereas, according to the standard model, after $e^{\pm}$ annihilation, the numbers of baryons and CMB photons in a comoving volume are unchanged. As a result, in the standard model, $\eta_{10}$ measured at present, at recombination, and at BBN are the same. Any departure would be a sign of new physics/cosmology beyond the standard models. This is one of the key cosmological tests. Because the baryon mass density ($\rho_{\rm B} \equiv \Omega_{\rm B}\rho_c$, where $\rho_c = 3H_0^2/8\pi G_N$ is the present critical mass density, $H_0$ is the present value of the Hubble parameter, and $G_N$ is the Newton constant) plays a direct role in the growth of perturbations, it is equally convenient to quantify the baryon abundance using a combination of $\Omega_{\rm B}$ and $h$, the present value of the Hubble parameter measured in units of 100 km s$^{-1}$ Mpc$^{-1}$,

$$\eta_{10} = 274\ \omega_{\rm B} \equiv 274\ \Omega_{\rm B}h^2. \qquad 2.$$

Until very recently, the comparison between the observationally inferred and the BBN-predicted primordial abundances, especially the relic abundance of D, provided the most accurate determination of $\eta_{10}$. However, this has now changed as a result of the very high-quality data from a variety of CMB experiments (2), complemented by observations of LSS (3). At present, the best non-BBN value (see References 2 and 3) is $\Omega_{\rm B}h^2 = 0.0223 \pm 0.0007$, corresponding to $\eta_{10} = 6.11 \pm 0.20$. This value is used below to predict the relic abundances for comparison with the observational data (and vice versa).

The Hubble parameter, $H = H(t)$, measures the expansion rate of the Universe. Deviations from the standard model ($H \rightarrow H'$) may be parameterized by an expansion rate parameter, $S \equiv H'/H$. In the standard model, during the early radiation-dominated evolution, $H$ is determined by the energy density in relativistic particles so that deviations from the standard cosmology ($S \neq 1$) may be quantified equally well by the "equivalent number of neutrinos," $\Delta N_\nu \equiv N_\nu - 3$:

$$\rho_{\rm R} \rightarrow \rho'_{\rm R} \equiv \rho_{\rm R} + \Delta N_\nu \rho_\nu. \qquad 3.$$

Prior to $e^{\pm}$ annihilation, these two parameters are related by

$$S = (1 + 7\Delta N_\nu/43)^{1/2}. \qquad 4.$$

$\Delta N_\nu$ and $S$ are equivalent ways to quantify any deviation from the standard model expansion rate; $\Delta N_\nu$ is not necessarily related to extra (or fewer) neutrinos. For example, if the value of the gravitational constant, $G_{\rm N}$, was different in the early Universe from its value at present, $S = (G'_{\rm N}/G_{\rm N})^{1/2}$, and $\Delta G_{\rm N}/G_{\rm N} = 7\Delta {\rm N}_\nu/43$ (4, 5). Changes (from the standard model) in the expansion rate at BBN will modify the neutron abundance and the time available for nuclear production/destruction, changing the BBN-predicted relic abundances.

Although most models of particle physics beyond the standard model adopt (or impose) a lepton asymmetry of the same order of magnitude as the (very small) baryon asymmetry ($\frac{n_B - n_{\bar{B}}}{n_\gamma} = 10^{-10}\eta_{10}$), lepton (neutrino) asymmetries that are orders of magnitude larger are currently not excluded by any experimental data. In analogy with $\eta_{10}$, which provides a measure of the baryon asymmetry, the lepton (neutrino) asymmetry, $L = L_\nu \equiv \Sigma_\alpha L_{\nu_\alpha}$ ($\alpha \equiv e, \mu, \tau$), may be quantified by the ratios of the neutral lepton chemical potentials to the temperature (in energy units) $\xi_{\nu_\alpha} \equiv \mu_{\nu_\alpha}/kT$, where

$$L_{\nu_\alpha} \equiv \left(\frac{n_{\nu_\alpha} - n_{\bar{\nu}_\alpha}}{n_\gamma}\right) = \frac{\pi^2}{12\zeta(3)}\left(\frac{T_{\nu_\alpha}}{T_\gamma}\right)^3 \xi_{\nu_\alpha}\left(1 + \left(\frac{\xi_{\nu_\alpha}}{\pi}\right)^2\right). \qquad 5.$$

Although any neutrino degeneracy always increases the energy density in the neutrinos, resulting in an effective $\Delta N_\nu > 0$, the range of $|\xi_{\nu_\alpha}|$ of interest to BBN is limited to sufficiently small values that the increase in $S$ arising from a nonzero $\xi_{\nu_\alpha}$ is negligible and $T_{\nu\alpha} = T_\gamma$ prior to $e^\pm$ annihilation (6). However, a small asymmetry between electron-type neutrinos and antineutrinos ($|\xi_e| \sim 10^{-2}$), although very large compared to the baryon asymmetry, can have a significant impact on BBN by modifying the pre-BBN n/p ratio. Because any charged lepton asymmetry is limited by the baryon asymmetry to be very small, any non-negligible lepton asymmetry must be among the neutral lepton, the neutrinos. For this reason, in the following, the subscript $\nu$ will be dropped ($\xi_{\nu_\alpha} \equiv \xi_\alpha$).

## 2.1. Big Bang Nucleosynthesis Chronology

When the Universe is only a fraction of a second old, at temperatures above several MeV, thermodynamic equilibrium has already been established among the key BBN players: neutrinos, $e^\pm$ pairs, photons, and nucleons. Consequently, without loss of generality, discussion of BBN can begin at this epoch. Early in this epoch, the charged-current weak interactions proceed sufficiently rapidly to keep the n/p ratio close to its equilibrium value $(n/p)_{eq} = e^{-\Delta m/T}$. Note that if there is an asymmetry between the numbers of $\nu_e$ and $\bar{\nu}_e$, the equilibrium n/p ratio is modified to $(n/p)_{eq} = \exp(-\Delta m/T - \mu_e/T) = e^{-\xi_e}(n/p)^0_{eq}$. As the Universe continues to expand and cool, the lighter protons are favored over the heavier neutrons and the n/p ratio decreases, initially tracking the equilibrium expression. However, as the temperature decreases below $T \sim 0.8$ MeV, when the Universe is ~1 s old, the weak interactions have become too slow to maintain equilibrium, and the n/p ratio, while continuing to fall, deviates from (exceeds) the equilibrium value (see **Figure 1**). Because the n/p ratio depends on the competition between the weak-interaction rates and the early-Universe expansion rate (as well as on a possible neutrino asymmetry), deviations from the standard model (e.g., $S \neq 1$ and/or $\xi_e \neq 0$) will change the relative numbers of neutrons and protons available for building the complex nuclides.

Simultaneously, nuclear reactions among the neutrons and protons (e.g., $n + p \leftrightarrow D + \gamma$) are proceeding very rapidly ($\Gamma_{\rm nuc} \gg H$). However, any D synthesized by this reaction finds itself bathed in a background of energetic $\gamma$ rays (the CMB blueshifted to average photon energies $E_\gamma \sim 3T_\gamma \gtrsim$ few MeV), and before another neutron or proton can be captured by the deuteron to begin building more complex nuclei, the

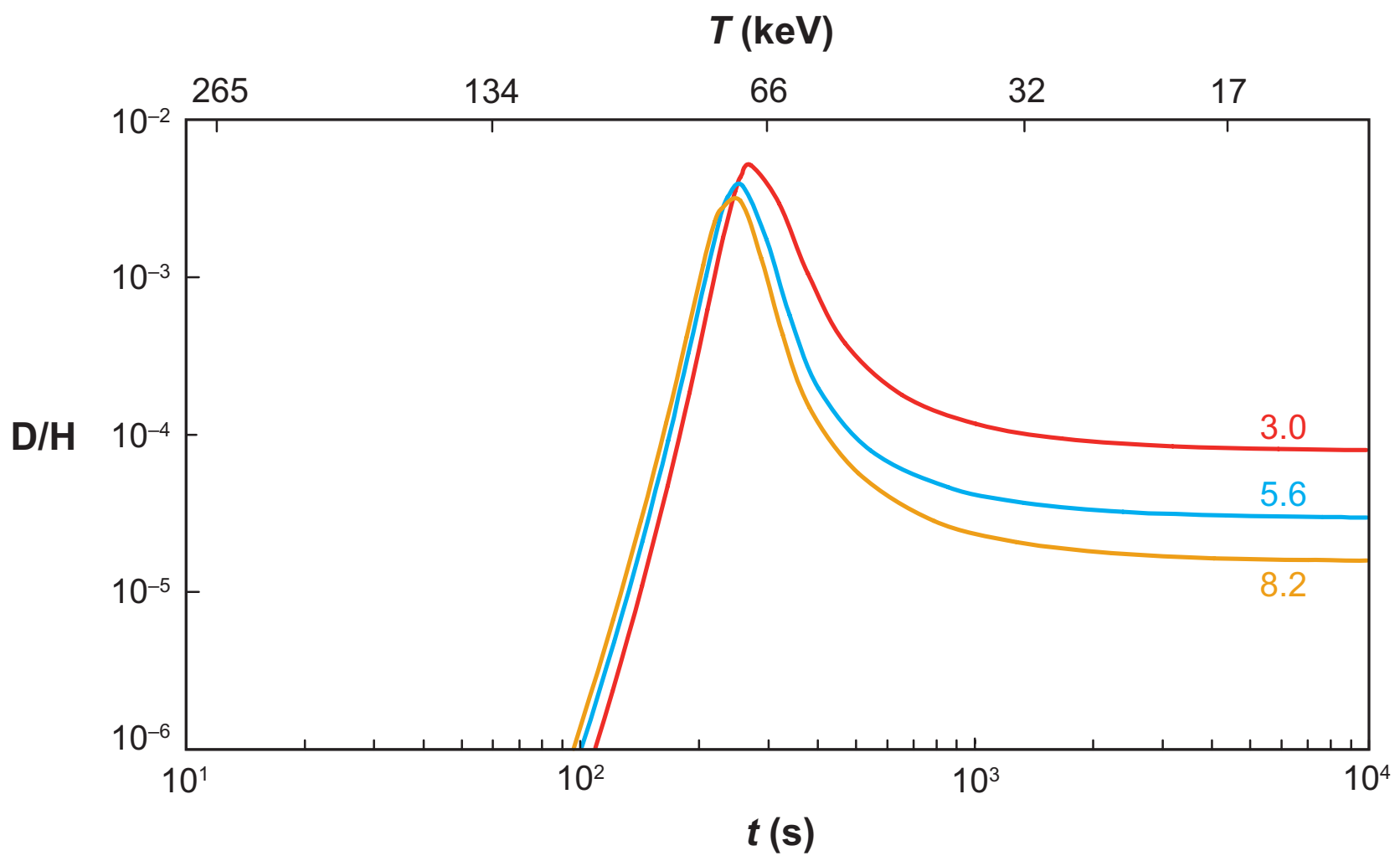

**Figure 2**

The time-temperature evolution of the deuteron abundance (the D/H ratio). The curves are labeled by the corresponding baryon abundances $\eta_{10}$.

deuteron is photodissociated and BBN remains stillborn. This bottleneck to BBN persists until the temperature drops even further below the deuteron binding energy, when there are too few sufficiently energetic photons to photodissociate the deuteron before it captures additional nucleons, launching BBN in earnest. This transition (smooth, but rapid) occurs after $e^{\pm}$ annihilation, when the Universe is now a few minutes old and the temperature has dropped below ~80 keV. In **Figure 2**, the time evolution of the ratio of deuterons to protons (D/H) is shown as a function of time; note the very rapid rise of the deuteron abundance. Once BBN begins, neutrons and protons quickly combine to build D, $^3$H, $^3$He, and $^4$He. The time evolution of the $^4$He (α-particle) abundance is shown in **Figure 3**; note that the rapid rise in $^4$He follows that in D. Because $^4$He is the most tightly bound of the light nuclides and

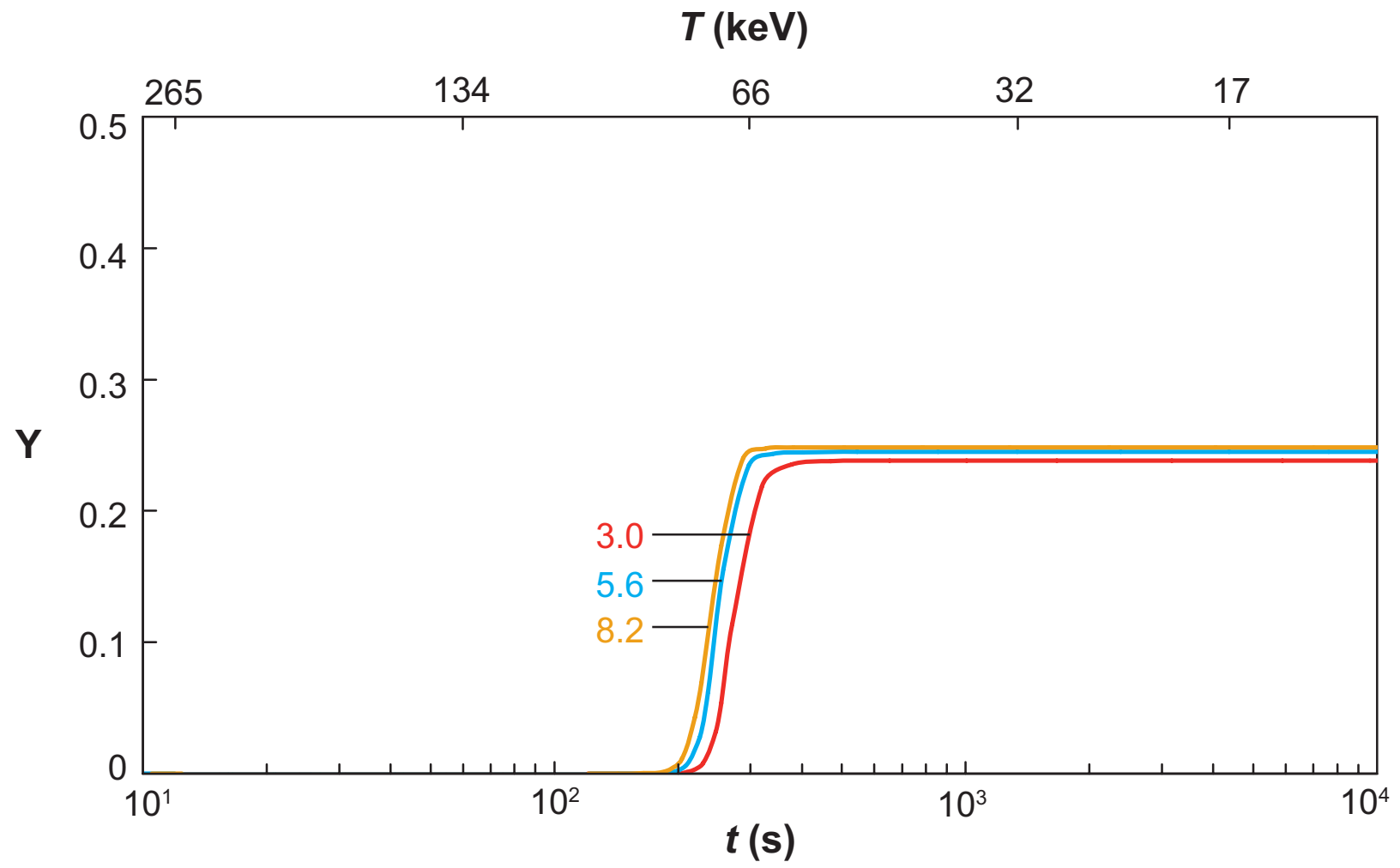

**Figure 3**

The time-temperature evolution of the $^4$He abundance $Y \equiv 4y/(1+4y)$, where $y \equiv n_{He}/n_H$. The curves are labeled by the corresponding baryon abundances $\eta_{10}$.

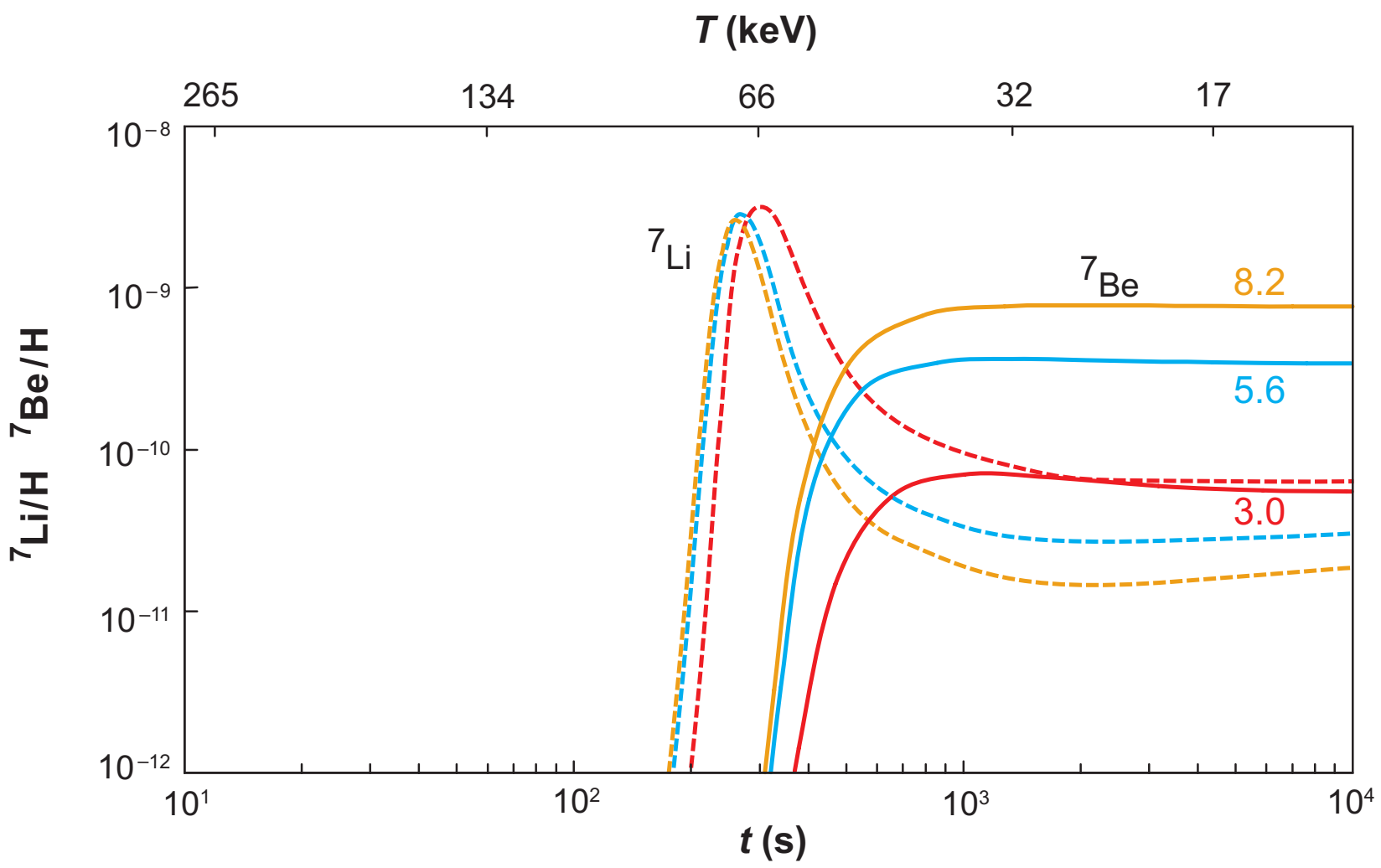

**Figure 4**

The time-temperature evolution of the mass-7 abundances. The solid curves are for direct production of $^7$Be, and the dashed curves are for direct production of $^7$Li. The curves are labeled by the corresponding baryon abundances, $\eta_{10}$.

because there are no stable mass-5 nuclides, a new bottleneck appears at $^4$He. As a result, the nuclear reactions quickly incorporate all available neutrons into $^4$He, and the $^4$He relic abundance is limited by the availability of neutrons at BBN. As may be seen in **Figure 3**, the $^4$He abundance is very insensitive to the baryon abundance.

To jump the gap at mass-5 requires Coulomb-suppressed reactions of $^4$He with D, $^3$H, or $^3$He, guaranteeing that the abundances of the heavier nuclides are severely depressed below that of $^4$He (and even that of D and $^3$He). The few reactions that do manage to bridge the mass-5 gap lead mainly to mass-7, producing $^7$Li and $^7$Be. Later, when the Universe has cooled further, $^7$Be captures an electron and decays to $^7$Li. As seen in **Figure 4**, direct production of $^7$Li by $^3$H($\alpha$, $\gamma$) $^7$Li reactions dominates at low baryon abundance ($\eta_{10} \lesssim 3$), whereas direct production of $^7$Be via $^3$He($\alpha$, $\gamma$) $^7$Be reactions dominates at higher baryon abundance ($\eta_{10} \gtrsim 3$). For the range of $\eta_{10}$ of interest, the BBN-predicted abundance of $^6$Li is more than three orders of magnitude below that of the more tightly bound $^7$Li. Finally, there is another gap at mass-8, ensuring that during BBN no heavier nuclides are produced in astrophysically interesting abundances.

The primordial nuclear reactor is short lived. As seen from **Figures 2–4**, as the temperature drops below $T \lesssim 30$ keV, when the Universe is ~20 min old, Coulomb barriers and the absence of free neutrons (almost all those present when BBN began have been incorporated into $^4$He) abruptly suppress all nuclear reactions. Afterward, until the first stars form, no relic primordial nuclides are destroyed (except that $^3$H and $^7$Be are unstable and decay to $^3$He and $^7$Li, respectively) and no new nuclides are created. In ~1000 s, BBN has run its course.

## 2.2. Standard Big Bang Nucleosynthesis–Predicted Abundances

For standard Big Bang Nucleosynthesis (SBBN), where $S = 1$ ($N_\nu = 3$) and $\xi_e = 0$, the BBN-predicted primordial abundances depend on only one free parameter, the

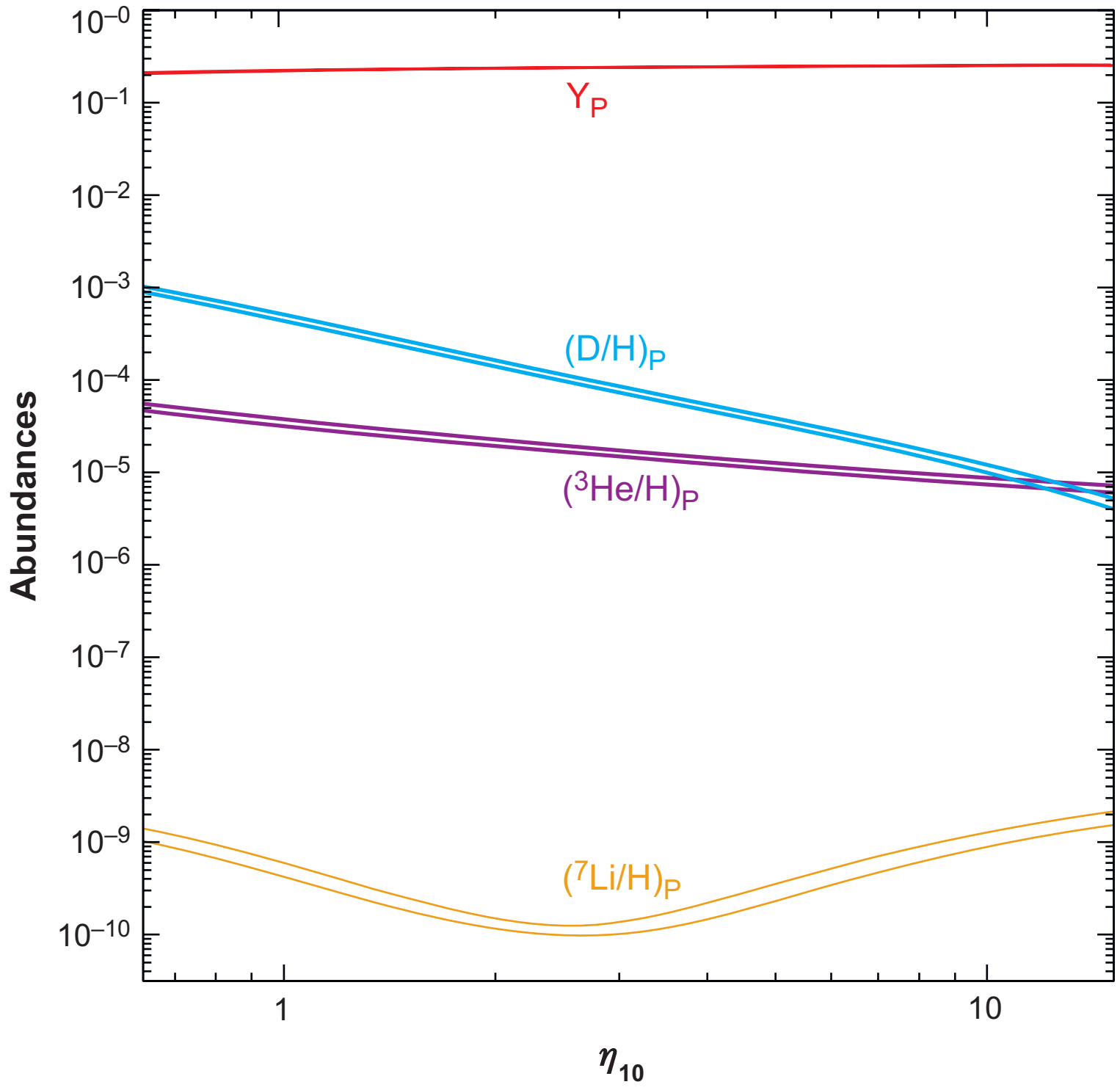

**Figure 5**

The SBBN-predicted primordial abundances of D, $^3$He, and $^7$Li (relative to hydrogen by number), and the $^4$He mass fraction ($Y_P$), as functions of the baryon abundance parameter $\eta_{10}$. The widths of the bands (including the band for $Y_P$!) reflect the uncertainties in the nuclear and weak-interaction rates.

baryon abundance $\eta_{10}$. In **Figure 5**, the SBBN-predicted relic abundances of D, $^3$He, $^4$He, and $^7$Li as a function of $\eta_{10}$ are shown. The trends revealed in **Figure 5** are easily understood on the basis of the preceding discussion. For example, D, $^3$H, and $^3$He are burned to $^4$He, and the higher the baryon abundance, the faster are D and $^3$He burned away and the smaller are their surviving abundances. Because the $^4$He abundance is limited by the abundance of neutrons, the primordial $^4$He mass fraction is very insensitive to $\eta_{10}$, $Y_P \equiv 4y/(1+4y) \approx \frac{2(n/p)_{BBN}}{1+(n/p)_{BBN}} \approx 1/4$ (see **Figure 3**), where $y \equiv n_{He}/n_H$. Of course, defined this way, $Y_P$ is not really the mass fraction because this expression adopts precisely 4 for the $^4$He-to-H mass ratio. However, the reader should be warned that $Y_P$ defined this way is conventionally referred to by cosmologists as the $^4$He mass fraction. The residual dependence of $Y_P$ on $\eta_{10}$ results from the fact that the higher the baryon abundance, the earlier the D bottleneck is breached—at a higher temperature, where the n/p ratio is slightly higher. As a result, $Y_P$ increases, but only logarithmically, with $\eta_{10}$. The valley shape of the $^7$Li abundance curve is a reflection of the two paths to mass-7 (see **Figure 4**). At low baryon abundance, the directly produced $^7$Li is burned away as the baryon abundance increases, whereas at higher baryon abundance, $^7$Be is synthesized more rapidly as the baryon abundance increases in the range of interest. Eventually, at much higher $\eta_{10}$, the $^7$Be will also be burned away as the baryon abundance increases.

Over the years, many have written computer codes to integrate the coupled set of differential equations that track element production/destruction to solve for the BBN-predicted abundances of the light nuclides. Once the cosmology is defined, the time-temperature-density relations are known and the only uncertain inputs are the nuclear and weak-interaction cross sections and rates. It is not surprising, indeed it is required, that with the same input the codes should predict the same abundances as a function of the one free parameter, the baryon abundance $\eta_{10}$. Now that observations of the CMB temperature fluctuations and of the distribution of LSS have become sufficiently precise, the range of $\eta_{10}$ of interest is considerably restricted (3): with $\sim$95% confidence, $5.7 \lesssim \eta_{10} \lesssim 6.5$. Within this limited range of $\eta_{10}$, there is no need to have access to a sophisticated, state-of-the-art BBN code, as the following simple fits are quite accurate (to within the nuclear and weak rate uncertainties) and are consistent with the published results of many independent BBN codes (7):

$$y_D \equiv 10^5(\mathrm{D/H}) = 2.68(1 \pm 0.03)(6/\eta_{10})^{1.6}, \quad 6.$$

$$y_3 \equiv 10^5(^3\mathrm{He/H}) = 1.06(1 \pm 0.03)(6/\eta_{10})^{0.6}, \quad 7.$$

$$\mathrm{Y_P} = 0.2483 \pm 0.0005 + 0.0016(\eta_{10} - 6), \quad 8.$$

$$y_{Li} \equiv 10^{10}(^7\mathrm{Li/H}) = 4.30(1 \pm 0.1)(\eta_{10}/6)^2. \quad 9.$$

Note that the lithium abundance is often expressed logarithmically as $[\mathrm{Li}] \equiv 12 + \log(\mathrm{Li/H}) = 2 + \log y_{Li}$. The expression for $\mathrm{Y_P}$ corresponds to the currently accepted value of the neutron lifetime, $\tau_n = 885.7 \pm 0.8$ s (8), which normalizes the strength of the charged-current weak interactions responsible for neutron-proton interconversions. Changes in the currently accepted value may be accounted for by adding $2 \times 10^{-4}(\tau_n - 885.7)$ to the right side of the expression for $\mathrm{Y_P}$. The above abundances have been calculated under the assumption that the three flavors of active neutrinos were populated in the pre-BBN Universe and had decoupled before $e^{\pm}$ annihilation, prior to BBN. Although this latter is a good approximation, it is not entirely accurate. Mangano et al. (9) relaxed the assumption of complete decoupling at the time of $e^{\pm}$ annihilation, finding that in the post-$e^{\pm}$ annihilation Universe (relevant for comparisons with the CMB and LSS) the relativistic energy density is modified (increased) in contrast to the complete decoupling limit. This can be accounted for by $\mathrm{N}_\nu$ (post-$e^{\pm}$ ann.) $= 3 \rightarrow 3.046$ (9). Mangano et al. find that the effect on the BBN yields is small, well within the uncertainties quoted above. For $^4$He, the predicted relic abundance (Equation 8) increases by $\sim 2 \times 10^{-4}$.

As an example to be explored more carefully below, if the CMB/LSS result for the baryon abundance is adopted ($\eta_{10} = 6.11 \pm 0.20$ at $\sim 1\sigma$), these analytic fits predict $y_D = 2.60 \pm 0.16$, $y_3 = 1.05 \pm 0.04$, $\mathrm{Y_P} = 0.2487 \pm 0.0006$, and $[\mathrm{Li}] = 2.65^{+0.05}_{-0.06}$. The corresponding 95% confidence ranges are $2.3 \lesssim y_D \lesssim 2.9$, $0.97 \lesssim y_3 \lesssim 1.12$, $0.247 \lesssim \mathrm{Y_P} \lesssim 0.250$, and $2.53 \lesssim [\mathrm{Li}] \lesssim 2.74$.

## 2.3. Nonstandard Big Bang Nucleosynthesis: $S \neq 1$

From the preceding description of BBN, it is straightforward to anticipate the changes in the BBN-predicted abundances when $S \neq 1$. Because D and $^3$He are being burned to $^4$He, a faster-than-standard expansion ($S > 1$) leaves less time for D and $^3$He destruction and their relic abundances increase, and vice versa for a slower-than-standard expansion. The $^4$He relic abundance is tied tightly to the neutron abundance at BBN. A faster expansion provides less time for neutrons to transform into protons, and the higher neutron fraction results in an increase in $Y_P$. The effect on the mass-7 abundance when $S \neq 1$ is a bit more complex. At low baryon abundance, $^7$Li is being destroyed and $S > 1$ inhibits its destruction, increasing the relic abundance of mass-7. In contrast, for high baryon abundance ($\eta_{10} \gtrsim 3$), direct production of $^7$Be dominates and a faster-than-standard expansion provides less time to produce mass-7, reducing its relic abundance. Isoabundance curves for D and $^4$He in the $S - \eta_{10}$ plane are shown in **Figure 6**. Although, in general, access to a BBN code is necessary to track the primordial abundances as functions of $\eta_{10}$ and $S$, Kneller & Steigman (10) have identified extremely simple but quite accurate analytic fits over a limited range in these variables ($4 \lesssim \eta_{10} \lesssim 8$, $0.85 \lesssim S \lesssim 1.15$, corresponding to $1.3 \lesssim N_\nu \lesssim 5.0$). For D and $^4$He [including the Mangano et al. (9) correction], these are

$$y_D \equiv 46.5(1 \pm 0.03)\eta_D^{-1.6}; \quad Y_P \equiv (0.2386 \pm 0.0006) + \eta_{He}/625, \qquad 10.$$

where

$$\eta_D \equiv \eta_{10} - 6(S - 1); \quad \eta_{He} \equiv \eta_{10} + 100(S - 1). \qquad 11.$$

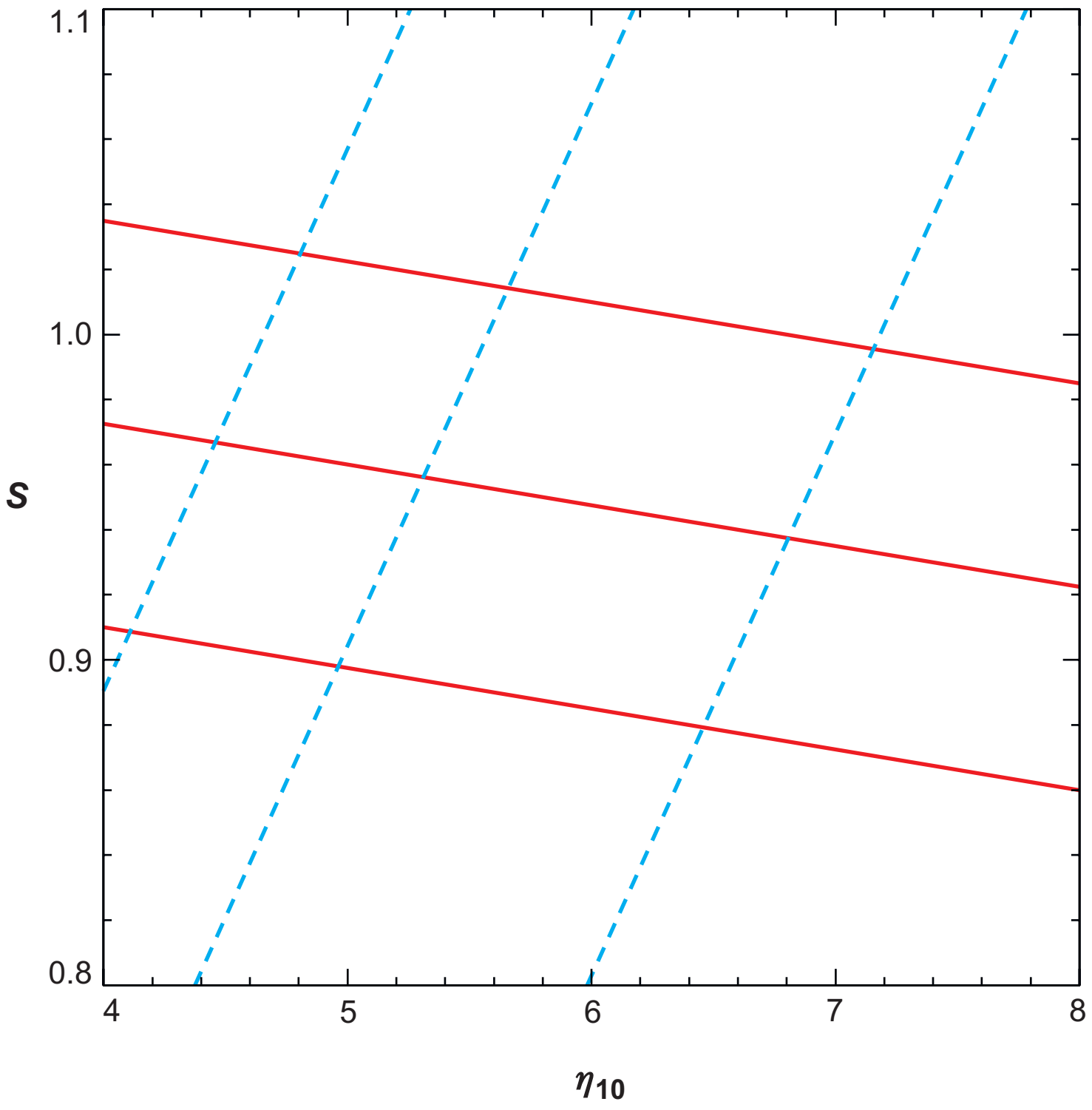

**Figure 6**

Isoabundance curves for D (*dashed blue lines*) and $^4$He (*solid red lines*) in the expansion rate factor ($S$)–baryon abundance ($\eta_{10}$) plane. The $^4$He curves, from bottom to top, are for $Y_P = 0.23, 0.24$, and $0.25$. The D curves, from left to right, are for $y_D = 4.0, 3.0$, and $2.0$.

Over the same high-$\eta_{10}$ range, the fit for $^7$Li is

$$y_{\rm Li} \equiv \frac{(1 \pm 0.1)}{8.5} \eta_{\rm Li}^2, \qquad 12.$$

where

$$\eta_{\rm Li} \equiv \eta_{10} - 3(S-1). \qquad 13.$$

Note that Kneller & Steigman (10) chose the coefficients in these fits to maximize the goodness of fit over the above ranges in $\eta_{10}$ and $S$, and not to minimize the difference between the fits and the more accurate results from the BBN code with $S = 1$. As a result, although these fits differ very slightly from those in Section 2.2, they do agree with them within the quoted $1\sigma$ errors.

As seen from **Figure 6** and the above equations, D (and $^7$Li) is (are) most sensitive to $\eta_{10}$, whereas $Y_{\rm P}$ is most sensitive to $S$. Observations that constrain the abundances of $^4$He and D (or $^7$Li) have the potential to constrain the cosmology/particle physics parameters $\eta_{10}$ and $S$. Note also that because $\eta_{\rm D}$ and $\eta_{\rm Li}$ depend very similarly on $\eta_{10}$ and $S$, constraining either the D or $^7$Li abundance has the potential to predict or constrain the other:

$$\eta_{\rm Li} = \eta_{\rm D} + 3(S-1) \approx \eta_{\rm D}. \qquad 14.$$

## 2.4. Nonstandard Big Bang Nucleosynthesis: $\xi_e \neq 0$

Although the most popular models for generating the baryon asymmetry in the Universe tie it to the lepton asymmetry, suggesting that $|L_{\nu_e}| \sim \xi_{\nu_e} \sim \eta_{10}$, this assumption is untested. An asymmetry in the electron-type neutrinos will modify the n/p ratio at BBN that affects the primordial abundance of $^4$He, with smaller effects on the other light-nuclide relic abundances. As already noted, for $\xi_{\nu_e} > 0$, the n/p ratio increases over its SBBN value, leading to an increase in the $^4$He yield. Isoabundance curves for D and $^4$He in the $\xi_e - \eta_{10}$ plane are shown in **Figure 7**. In this case too, Kneller & Steigman (10) identified simple analytic fits that are quite accurate over limited ranges in $\eta_{10}$ and $\xi_e$ ($4 \lesssim \eta_{10} \lesssim 8$, $-0.1 \lesssim \xi_e \lesssim 0.1$):

$$\eta_{\rm D} \equiv \left(\frac{46.5(1 \pm 0.03)}{y_{\rm D}}\right)^{1/1.6} = \eta_{10} + 5\xi_e/4, \qquad 15.$$

$$\eta_{\rm He} \equiv 625(Y_{\rm P} - 0.2386 \pm 0.0006) = \eta_{10} - 574\xi_e/4, \qquad 16.$$

$$\eta_{\rm Li} \equiv (8.5(1 \pm 0.1) y_{\rm Li})^{1/2} = \eta_{10} - 7\xi_e/4. \qquad 17.$$

As for the case of a nonstandard expansion rate, here the $^4$He abundance provides the most sensitive probe of lepton asymmetry, whereas D and $^7$Li mainly constrain the baryon abundance and their abundances are strongly correlated:

$$\eta_{\rm Li} = \eta_{\rm D} - 3\xi_e \approx \eta_{\rm D}. \qquad 18.$$

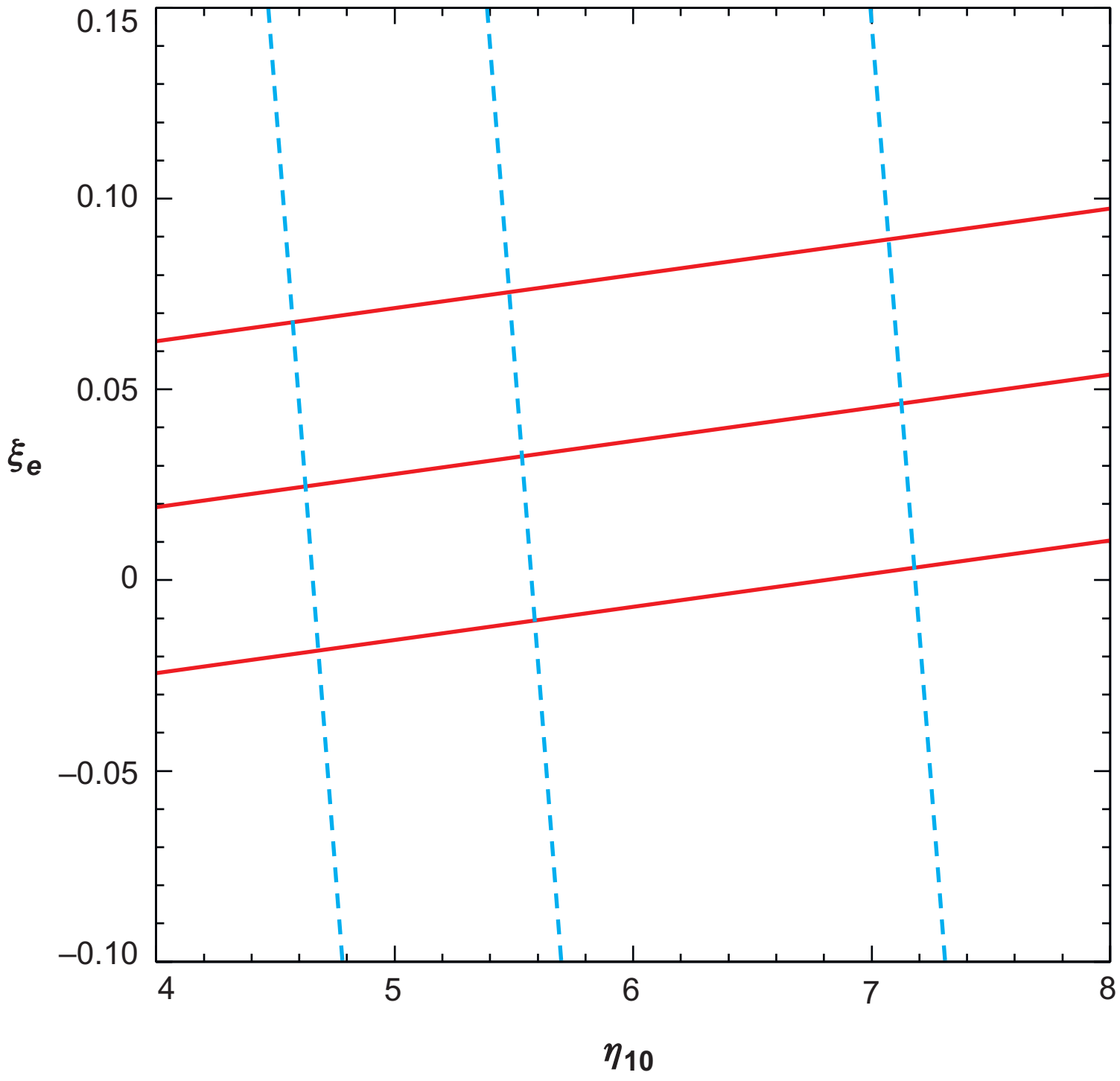

**Figure 7**

Isoabundance curves for D (*dashed blue lines*) and $^4$He (*solid red lines*) in the lepton asymmetry ($\xi_e$)–baryon abundance ($\eta_{10}$) plane. The $^4$He curves, from top to bottom, are for $Y_P = 0.23$, 0.24, and 0.25. The D curves, from left to right, are for $y_D = 4.0$, 3.0, and 2.0.

### 2.5. Nonstandard Big Bang Nucleosynthesis: $S \neq 1$ and $\xi_e \neq 0$

The two cases of nonstandard BBN discussed in the preceding sections may be combined using the linear relations among the three parameters derived from observations of the relic abundances ($\eta_D$, $\eta_{He}$, $\eta_{Li}$) and the three cosmological/particle physics parameters ($\eta_{10}$, $S$, $\xi_e$). However, although in principle observations of the relic abundances of D, $^4$He, and $^7$Li can constrain $\eta_{10}$, $S$, and $\xi_e$, in practice the observational uncertainties are too large at present to lead to useful constraints. Therefore, the strategy in the following discussion is to use the inferred relic abundances of D and $^4$He to constrain the pairs $\{\eta_{10}, S\}$ or $\{\eta_{10}, \xi_e\}$ and to use the results to predict the BBN abundance of $^7$Li.

## 3. THE RELIC NUCLIDES OBSERVED

In the precision era of cosmology, systematic errors have come to dominate over statistical uncertainties. This is true, with a vengeance, of the relic abundances of the BBN-produced light nuclides inferred from current observational data. Although the bad news is that these largely uncontrolled systematic errors affect D, $^3$He, $^4$He, and $^7$Li, the good news is that each of these nuclides is observed in completely different astronomical objects by entirely different observational techniques so that these errors are uncorrelated. Furthermore, the post-BBN evolution of these nuclides, with its potential to modify the relic abundances, is different for each of them.

For example, owing to its very weak binding, any D incorporated into stars is completely burned away during the collapse of the prestellar nebula. For the same reason, any D synthesized in stellar interiors is burned away, to $^3$He and beyond, before it can be returned to the interstellar medium (ISM). As a result, the post-BBN evolution of D is simple and monotonic: It is only destroyed, with the consequence being that D observed anywhere at any time in the post-BBN evolution of the Universe provides a lower bound to the primordial D abundance (11). In contrast, the post-BBN evolution of the more tightly bound $^3$He nucleus is more complex. Prestellar D is burned to $^3$He, which, along with any prestellar $^3$He, may be preserved in the cooler outer layers of stars or burned away in the hotter interiors. In addition, hydrogen burning in the relatively cooler interiors of lower-mass stars may result in the net production of new $^3$He. If this material survives the later stages of stellar evolution, it may result in stellar-produced $^3$He being returned to the ISM. The bottom line is that stellar and galactic evolution models are necessary to track the post-BBN evolution of $^3$He in regions containing stellar-processed material, with the result being that the relic abundance inferred from such observations is model dependent (12).

As gas is cycled through successive generations of stars, hydrogen is burned to helium ($^4$He) and beyond, with the net effect being that the post-BBN abundance of $^4$He increases along with the abundances of the heavier elements, "metals" such as C, N, and O. This contamination of the relic $^4$He abundance by stellar-produced helium is non-negligible, although by tracking its dependence on the metallicity (e.g., oxygen abundance) observed in the same astrophysical sites provides a means to estimate and correct for such pollution. In general, however, the extrapolation to zero metallicity introduces uncertainties in the inferred $^4$He primordial abundance.

Finally, although $^7$Li is, like D, a very weakly bound nucleus, its post-BBN evolution is more complicated than that of D. Whereas any $^7$Li in the interiors of stars is burned away, some $^7$Li in the cooler outer layers of the lowest-mass, coolest stars, where the majority of the spectra used to infer the lithium abundances are formed, may survive. Furthermore, observations of enhanced lithium in some evolved stars (super-lithium-rich red giants) suggest that $^7$Li formed in the hotter interiors of some stars may be transported to the cooler exteriors before being destroyed. If this lithium-rich material is returned to the ISM before being burned away, these (and other) stars may be net lithium producers. In addition, $^7$Li is synthesized (along with $^6$Li, $^9$Be, and $^{10,11}$B) in nonstellar processes involving collisions of cosmic ray nuclei (protons, alphas, and CNO nuclei) with their counterparts in the ISM. Because stellar-synthesized CNO nuclei are necessary for spallation-produced $^7$Li, this provides a correlation between $^7$Li and the metallicity, which may help track this component of post-BBN lithium synthesis. However, production of $^7$Li via $\alpha$-$\alpha$ fusion may occur prior to the production of the heavier elements, mimicking relic production (13).

Below, the observations relevant to inferring the BBN abundances of each of these nuclides are critically reviewed. We emphasize the distinction between statistical precision and the limited accuracy—which results from difficult-to-quantify systematic uncertainties in the observational data, within the context of our understanding of the post-BBN evolution of these relic nuclei and of the astrophysical sites where they are observed.

## 3.1. Deuterium: The Baryometer of Choice

Deuterium's simple post-BBN evolution, combined with its sensitivity to the baryon abundance ($y_D \propto \eta_{10}^{-1.6}$), singles it out among the relic nuclides as the baryometer of choice. Although there are observations of D in the solar system (14) and the ISM of the Galaxy (15), which provide interesting lower bounds to its primordial abundance, any attempt to employ those data to infer the primordial abundance introduces model-dependent, galactic evolution uncertainties. From the discussion above, the D abundance is expected to approach its BBN value in systems of very low metallicity and/or those observed at earlier epochs (high redshift) in the evolution of the Universe. As the metallicity ($Z$) decreases and/or the redshift ($z$) increases, the corresponding D abundances should approach a plateau at the BBN-predicted abundance. Access to D at high $z$ and low $Z$ is provided by observations of the absorption by neutral gas of light emitted by distant QSOs. By comparing observations of absorption due to D with the much larger absorption by hydrogen in these QSO absorption-line systems (QSOALS), the nearly primordial D/H ratio may be inferred.

The identical absorption spectra of DI and HI (modulo the ~81 km s$^{-1}$ velocity offset that results from the isotope shift) are a severe liability, creating the potential for confusion of DI absorption with that of an HI interloper masquerading as DI (16). This is exacerbated by the fact that there are many more HI absorbers at low column density than at high column density. As a result, it is necessary to select very carefully those QSOALS with simple, well-understood velocity structures. This selection process is telescope intensive, leading to the rejection of D/H determinations of many potential QSOALS targets identified from low-resolution spectra, after having invested the time and effort to obtain the necessary high-resolution spectra. This has drastically limited the number of useful targets in the otherwise vast Lyman-α forest of QSO absorption spectra (see Reference 17 for further discussion).

The higher HI column-density absorbers (e.g., damped Lyman-α absorbers) have advantages over the lower HI column-density absorbers (Lyman limit systems) by enabling observations of many lines in the Lyman series. However, a precise determination of the damped Lyman-α HI column density using the damping wings of the HI absorption requires an accurate placement of the continuum, which could be compromised by HI interlopers, leading to the potential for systematic errors in the inferred HI column density. These complications are real, and the path to primordial D using QSOALS has been fraught with obstacles, with some abundance claims having had to be withdrawn or revised. As a result, despite much work utilizing some of the largest telescopes, through 2006 there have been only six sufficiently simple QSOALS with D detections that lead to reasonably robust abundance determinations (see Reference 18 and further references therein). These are shown in **Figure 8**, in addition to, for comparison, the corresponding solar system and ISM D abundances. It is clear from **Figure 8** that there is significant dispersion, in excess of the claimed observational errors, among the derived D abundances at low metallicity, which, so far, masks the anticipated primordial D plateau. This large dispersion suggests that systematic errors, whose magnitudes are hard to estimate, may have contaminated the determinations of (at least some of) the DI and/or HI column densities.

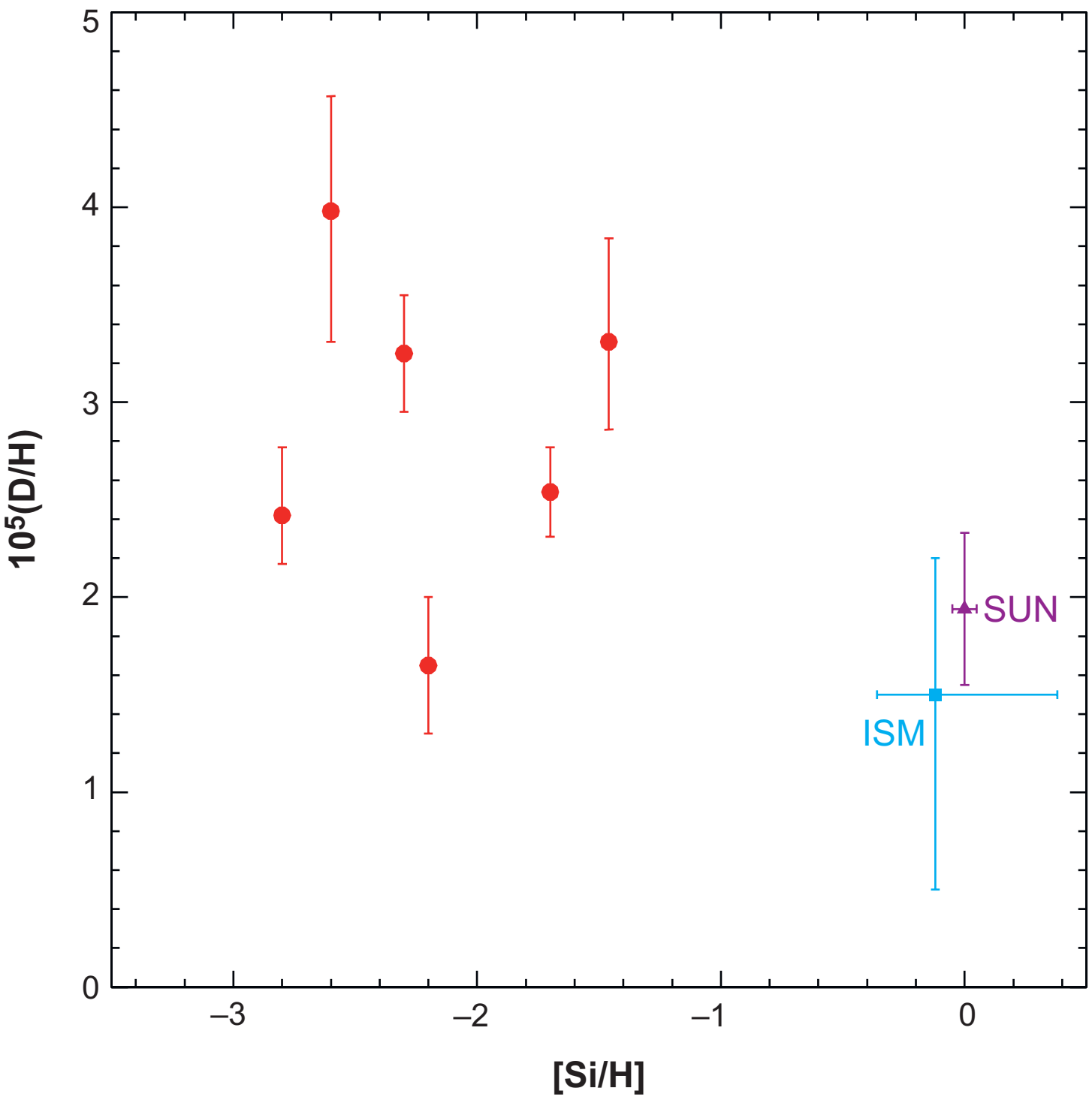

**Figure 8**

Deuterium abundances derived from observations of high-redshift, low-metallicity QSOALS as a function of the corresponding metallicities (shown relative to solar on a log scale). Also shown for comparison are the D abundances derived from solar system observations (SUN), as well as the range in D and oxygen abundances inferred from observations of the ISM.

O'Meara et al. (18) choose the mean of the log of the six individual D abundances as an estimator of the primordial D abundance, finding $\log(y_{DP}) = 0.454 \pm 0.036$. This choice, corresponding to $y_{DP} = 2.84^{+0.25}_{-0.23}$, has $\chi^2 = 18.5$ for five degrees of freedom. This author is unconvinced that the mean of the log of the individual D abundances is the best estimator of the relic D abundance. If, instead, the $\chi^2$ of the individual $y_D$ determinations is minimized, $\langle y_D \rangle = 2.68$ and $\chi^2_{min} = 19.3$ (for five dof). Although the $\chi^2$ for either of these estimators is excessive, we prefer to use the individual $y_D$ determinations to find $\langle y_D \rangle$. In an attempt to compensate for the large dispersion among the individual $y_D$ values, the individual errors are inflated by a factor of $(\chi^2_{min}/\mathrm{dof})^{1/2} = 1.96$, which leads to the following estimate of the primordial D abundance that is adopted for the subsequent discussion:

$$y_{DP} = 2.68^{+0.27}_{-0.25}. \qquad 19.$$

## 3.2. Helium-3

In contrast to D, which is observed in neutral gas via absorption, $^3$He is observed in emission from regions of ionized gas, HII regions. The $^3$He nucleus has a net spin so that for singly ionized $^3$He, the analog of the 21-cm spin-flip transition in neutral hydrogen occurs at 3.46 cm, providing the $^3$He observational signature. The

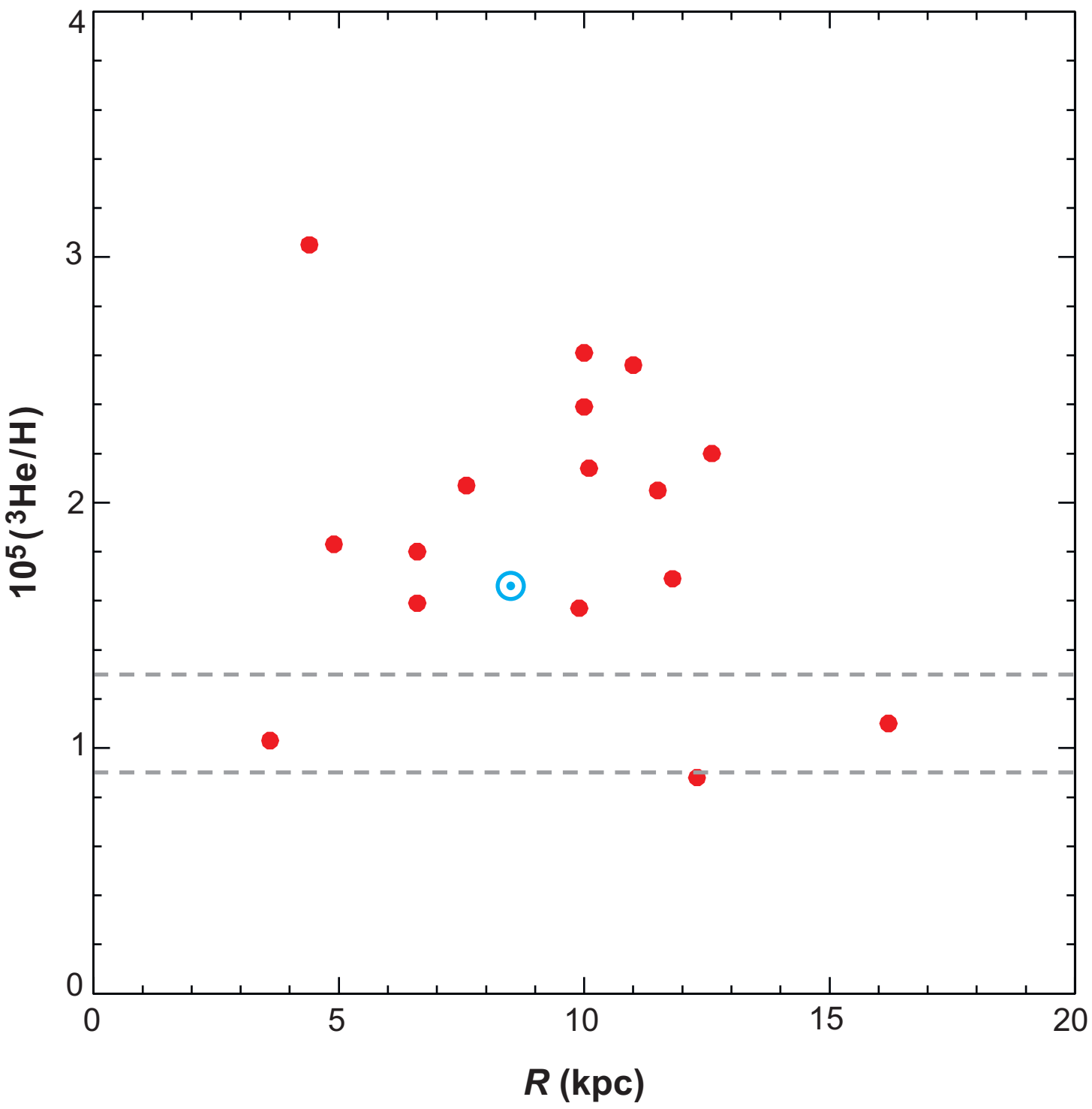

**Figure 9**

The $^3$He abundances (by number relative to hydrogen) derived from observations of HII regions in the Galaxy (19) as a function of the corresponding distances from the galactic center (*R*). The blue solar symbol indicates the $^3$He abundance for the presolar nebula (14). The dashed gray lines show the 1$\sigma$ band adopted by Bania et al. (19) for an upper limit to the primordial $^3$He abundance.

emission is quite weak, with the result that $^3$He is observed only (outside of the solar system) in HII regions and in a few planetary nebulae in the Galaxy. The latter confirm the net stellar production of $^3$He in at least some stars. In the Galaxy, there is a clear gradient of metallicity with distance from the center, with higher metallicity—more stellar processing—in the center and less in the suburbs. If in the course of the chemical evolution of the Galaxy, the abundance of $^3$He increases (net production) or decreases (net destruction), a gradient in $^3$He abundance (with metallicity and/or with distance) is also expected. However, as is clear from **Figure 9**, the data (14, 19) reveal no statistically significant correlation between the $^3$He abundance and location (or metallicity) in the Galaxy. This suggests a very delicate balance between post-BBN net production and net destruction of $^3$He. For a review of the current status of $^3$He evolution, see Romano et al. (20). Although the absence of a gradient suggests that the mean $^3$He abundance in the Galaxy ($y_3 \approx 1.9 \pm 0.6$) may provide a good estimate of the primordial abundance, Bania et al. (19) instead prefer to adopt as an upper limit to the primordial abundance the $^3$He abundance inferred from observations of the most distant (from the galactic center), most metal-poor, galactic HII region, $y_{3P} \lesssim 1.1 \pm 0.2$ (see **Figure 9**). For purposes of this review, the Bania et al. (19) value is adopted as an estimate of the primordial abundance of $^3$He.

$$y_{3P} = 1.1 \pm 0.2. \qquad 20.$$

Given that the post-BBN evolution of $^3$He, involving competition among stellar production, destruction, and survival, is considerably more complex and model dependent than that of D, and that observations of $^3$He are restricted to the chemically evolved solar system and the Galaxy, the utility of $^3$He as a baryometer is limited. In the subsequent comparison of predictions and observations, $^3$He will mainly be employed to provide a consistency check.

### 3.3. Helium-4

The post-BBN evolution of $^4$He is quite simple. As gas cycles through successive generations of stars, hydrogen is burned to $^4$He (and beyond), increasing the $^4$He abundance above its primordial value. The $^4$He mass fraction in the present Universe, $Y_0$, has received significant contributions from post-BBN stellar nucleosynthesis, so that $Y_0 > Y_P$. However, because some elements such as oxygen are produced by short-lived massive stars and $^4$He is synthesized (to a greater or lesser extent) by all stars, at very low metallicity the increase in Y should lag that in O/H, resulting in a $^4$He plateau, with $Y \to Y_P$ as $O/H \to 0$. Therefore, although $^4$He is observed in the Sun and in galactic HII regions, to minimize model-dependent evolutionary corrections, the key data for inferring its primordial abundance are provided by observations of helium and hydrogen emission lines generated from the recombination of ionized hydrogen and helium in low-metallicity extragalactic HII regions. The present inventory of such regions studied for their helium content exceeds 80 [see Izotov & Thuan (IT) (21)]. Because for such a large data set even modest observational errors can result in an inferred primordial abundance whose formal statistical uncertainty may be quite small, special care must be taken to include hitherto ignored or unaccounted for systematic corrections. It is the general consensus that the present uncertainty in $Y_P$ is dominated by the latter, rather than by the former, errors.

Although astronomers have generally been long aware of important sources of potential systematic errors (22), attempts to account for them have often been unsystematic or entirely absent. When it comes to using published estimates of $Y_P$, *caveat emptor*. The current conventional wisdom that the accuracy of the data demands the inclusion of systematic errors has led to recent attempts to account for some of them (21–28). In **Figure 10**, a sample of $Y_P$ determinations from 1992 to 2006 is shown (29). Most of these observationally inferred estimates (largely uncorrected for systematic errors or corrected unsystematically for some) fall below the SBBN-predicted primordial abundance (see Section 2.2), hinting either at new physics or at the need to account more carefully and consistently for all known systematic corrections.

Keeping in mind that more data do not necessarily translate to higher accuracy, the largest observed, reduced, and consistently analyzed data set of helium abundance determinations from low-metallicity extragalactic HII regions is that from IT 2004 (21). For their full sample of more than 80 HII regions, IT infer $Y_P^{IT} = 0.243 \pm 0.001$. From an analysis of an a posteriori–selected subset of seven of these HII regions, in which they attempt to account for some of the systematic errors, IT derive a consistent, slightly smaller value of $Y_P = 0.242 \pm 0.002$. Both of these $Y_P$ values are shown in **Figure 10**. The IT analysis of this subset is typical of many of the recent attempts

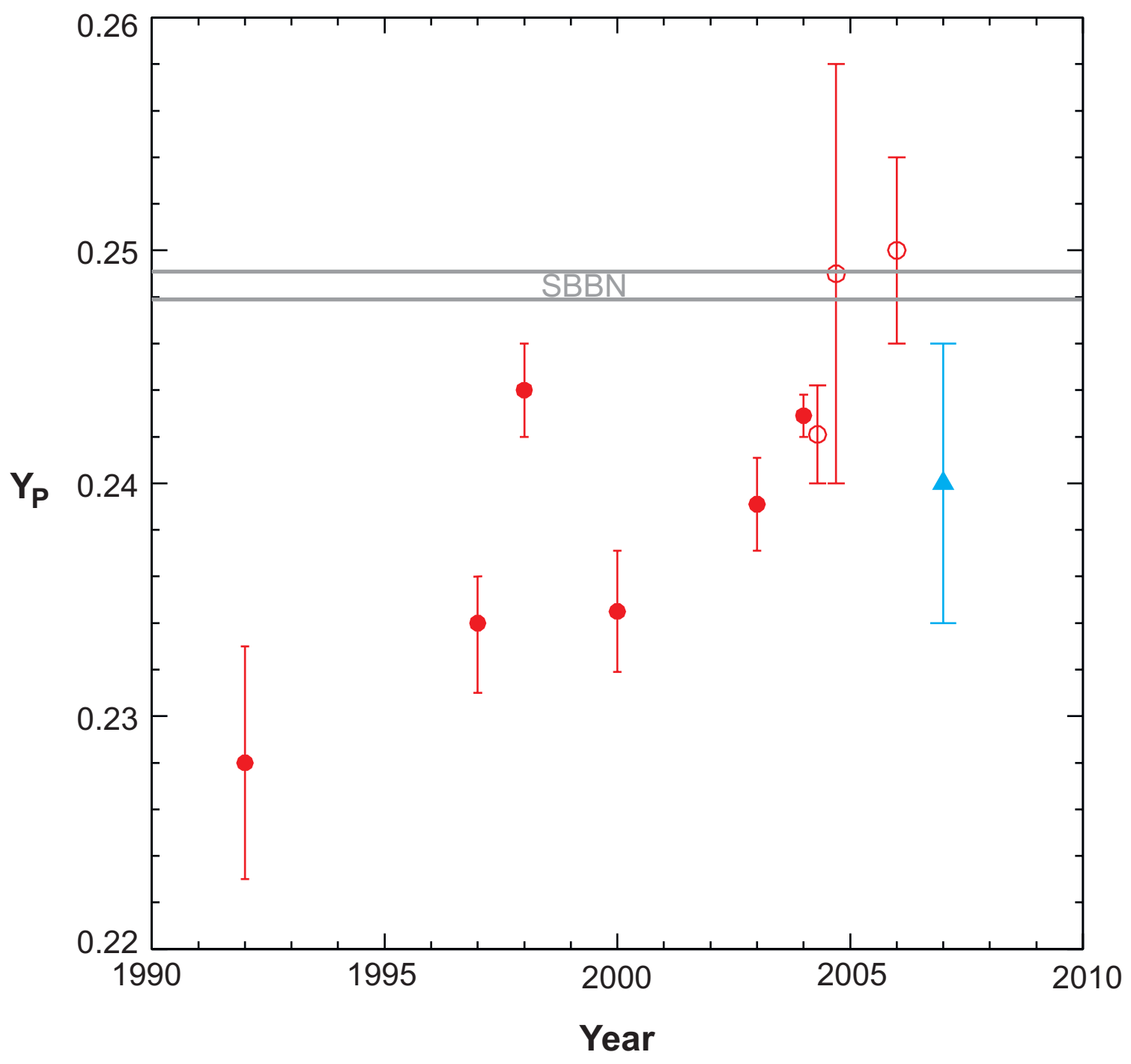

**Figure 10**

A sample of the observationally inferred primordial $^4$He abundances published from 1992 to 2006 (see text). The error bars are the quoted 1$\sigma$ uncertainties. The open red circles are derived from various a posteriori–selected subsets of the IT 2004 HII regions. The filled blue triangle is the value of $Y_P$ adopted in this review. Also shown is the 1$\sigma$ band for the SBBN-predicted relic abundance.

to include systematic corrections: Almost always only a subset of the known sources of systematic errors is analyzed, and almost always these analyses are applied to a very limited set (typically 1–7) of HII regions, which have usually been selected a posteriori.

For example, Luridiana et al. (24) use photoionization models to account for the effect of collisional excitation of Balmer lines for three (out of a total of five) HII regions, and extrapolate their results for the individual helium and oxygen abundances to zero oxygen abundance using the slope of the Y-versus-O relation derived from chemical evolution models. Their result, $Y_P = 0.239 \pm 0.002$, is shown in **Figure 10**.

The Olive & Skillman (OS) (23) and Fukugita & Kawasaki (FK) (27) analyses of the IT data attempt to account for the effect of underlying stellar absorption on the helium and hydrogen emission lines. Following criteria outlined in their 2001 paper (23), OS found they could apply their analysis to only 7 of the 82 IT HII regions. This small data set, combined with its restricted range in metallicity (oxygen abundance), severely limits the statistical significance of the OS conclusions. For example, there is no evidence from the seven OS HII regions that $\Delta Y \equiv Y^{OS} - Y^{IT}$ is correlated with metallicity, and the weighted mean correction and the error in its mean are $\Delta Y = 0.0029 \pm 0.0032$ (the average correction and its average error are $\Delta Y = 0.0009 \pm 0.0095$), consistent with zero at 1$\sigma$. If the weighted mean offset is applied to the IT-derived primordial abundance of $Y_P^{IT} = 0.243 \pm 0.001$, the corrected primordial value is $Y_P^{IT/OS} = 0.246 \pm 0.004$ (where, to be conservative, the errors have been added linearly). In contrast, OS prefer to force a fit of the seven data points to a linear Y-versus-O/H relation and from it derive the primordial abundance.

It is not surprising that for only seven data points spanning a relatively narrow range in metallicity, their linear fit, $Y_7^{OS} = 0.2495 \pm 0.0092 + (54 \pm 187)$ (O/H), is not statistically significant. In particular, the large error in $Y_P$ is dominated by the very large uncertainty in the slope of the Y-versus-O relation (which includes unphysical, negative slopes). Indeed, the OS linear fit is not preferred statistically over the simple weighted mean of the seven OS helium abundances because the reduced $\chi^2$ ($\chi^2$/dof) is actually higher for the linear fit. All eight HII regions reanalyzed by OS are consistent with a weighted mean, along with the error in the mean, of $\langle Y \rangle = 0.250 \pm 0.002$ ($\chi^2$/dof $= 0.51$). This result is of interest in that it provides an upper bound to primordial helium, $Y_P \leq \langle Y \rangle \leq 0.254$ at $\sim 2\sigma$.

Fukugita & Kawasaki (27) performed a very similar analysis to that of OS using 30 of the IT HII regions and the Small Magellanic Cloud HII region from Peimbert et al. (24). For this larger data set, Fukugita & Kawasaki (27) found an anticorrelation between their correction, $\Delta Y$, and the oxygen abundance. This flattens the FK-inferred Y-versus-O/H relation to the extent that they too found no evidence for a statistically significant correlation of He and oxygen abundances ($dY/dZ = 1.1 \pm 1.4$), leading to a zero-metallicity intercept, $Y_P^{FK} = 0.250 \pm 0.004$. As with the OS analysis, it seems that the most robust conclusion that can be drawn from the FK analysis is to use the weighted mean, along with the error in the mean of their 31 HII regions ($\chi^2$/dof $=$ 0.58) to provide an upper bound to primordial helium: $Y_P < \langle Y \rangle = 0.253 \pm 0.001 \leq 0.255$ at $\sim 2\sigma$.

Very recently, Peimbert, Luridiana, & Peimbert (PLP07) (28), using new atomic data, reanalyzed five HII regions spanning a factor of six in metallicity. Four of their five HII regions were in common with those analyzed by OS. PLP07, too, found no support from the data for a positive correlation between helium and oxygen. In **Figure 11**, the OS and PLP07 results for He/H and O/H (the ratios by number for helium to hydrogen and for oxygen to hydrogen, respectively) are shown. Although PLP07 force a model-dependent linear correlation to their data that, when extrapolated to zero metallicity, leads to $Y_P = 0.247 \pm 0.003$, their data are better fit by the weighted mean ($\chi^2$/dof $= 0.07$) and the error in the mean: $\langle Y \rangle = 0.251 \pm 0.002$, leading to an upper bound to $Y_P < \langle Y \rangle \leq 0.255$ at $\sim 2\sigma$.

There are other sources of systematic errors that have not been included in (some of) these and other analyses. For example, because hydrogen and helium recombination lines are used, the observations are blind to any neutral helium or hydrogen. Estimates of the ionization correction factor (icf), although model dependent, are large (26). Using models of HII regions ionized by the distribution of stars of different masses and ages and comparing them to the IT 1998 data, Gruenwald et al. (GSV) (26) concluded that the IT analysis overestimated the primordial $^4$He abundance by $\Delta Y^{GSV}(\text{icf}) \approx 0.006 \pm 0.002$; Sauer & Jedamzik (26) found an even larger correction. If the GSV correction is applied to the OS-revised, IT primordial abundance, the icf-corrected value becomes $Y_P^{IT/OS/GSV} = 0.240 \pm 0.006$ (as above, the errors have been added linearly).

The lesson from our discussion (and from **Figure 10**) is that although recent attempts to determine the primordial abundance of $^4$He may have achieved high precision, their accuracy remains in question. The latter is limited by our understanding

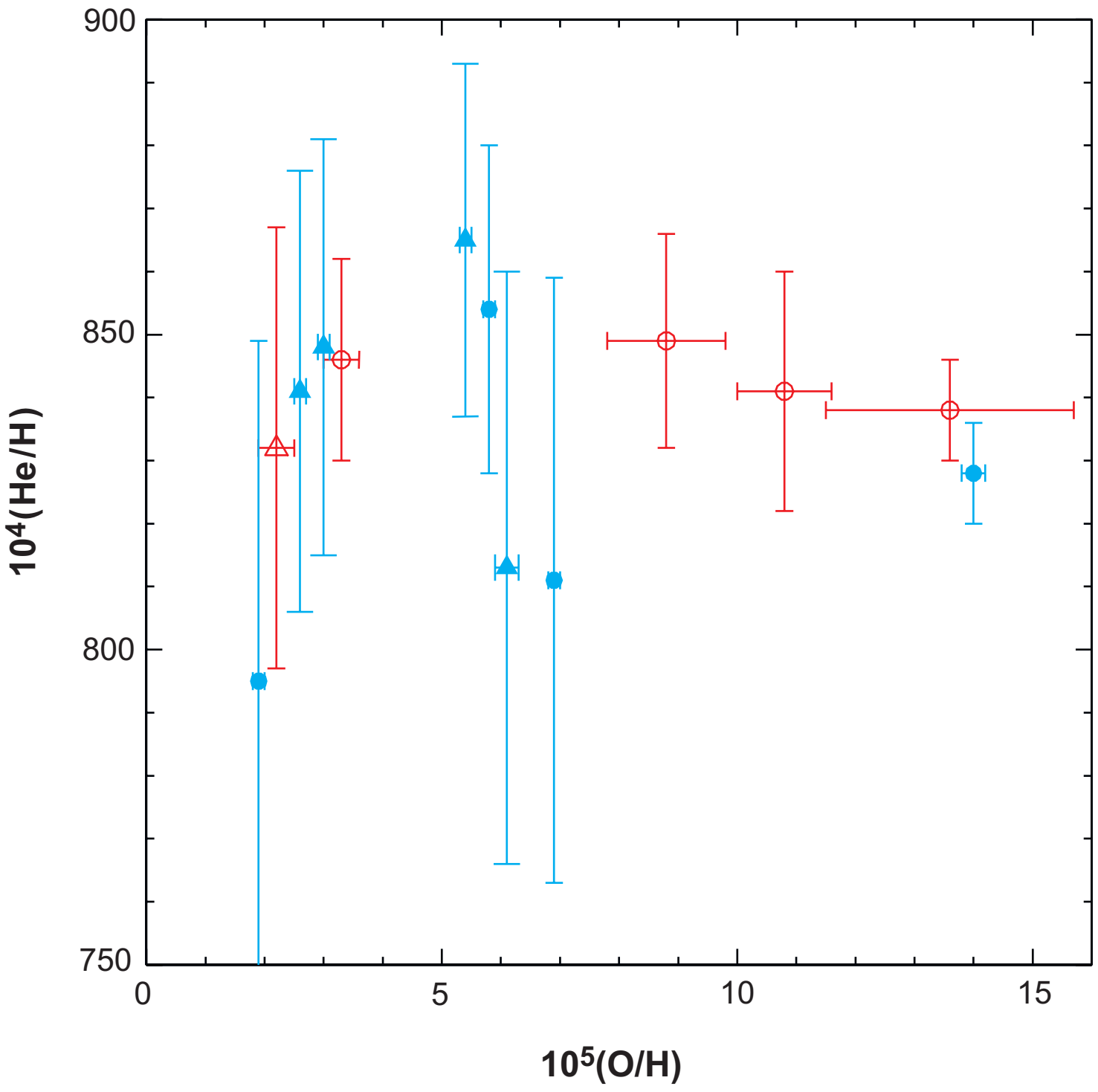

**Figure 11**

The HII region helium-to-hydrogen (He/H) ratios (by number) as a function of the oxygen-to-hydrogen (O/H) ratios (by number) from the analyses by Olive & Skillman (*blue filled triangles and circles*) (23) and by Peimbert et al. (*red open triangles and circles*) (28). The filled and open circles are for the HII regions common to the two analyses.

of and ability to account for systematic corrections and their errors, not by the statistical uncertainties. The good news is that carefully planned and targeted studies of the lowest-metallicity extragalactic HII regions may go a long way toward a truly accurate determination of $Y_P$. In the best of all worlds, a team of astronomers would develop a well-defined observing strategy designed to acquire the data necessary to address all the known sources of systematic error, identify an a priori–selected set of target HII regions, and carry out the observations and analyses. The bad news is that too few astronomers and telescope time-allocation committee members are either aware or convinced that this is interesting and important and worth their effort and dedicated telescope time.

In the opinion of this author, the question of the observationally determined value of $Y_P$ (and its error) is currently unresolved. The most recent analyses (OS, FK, PLP07) fail to find evidence for the anticipated correlation between the helium and oxygen abundances, calling into question the model-dependent extrapolations to zero metallicity often employed in the quest for primordial helium. Perhaps the best that can be done at present is to adopt a defensible value for $Y_P$ and, especially, its uncertainty. To this end, the $Y_P^{IT/OS/GSV}$ estimate is chosen for the subsequent discussion:

$$Y_P = 0.240 \pm 0.006. \qquad 21.$$

The adopted error is an attempt to account for the systematic as well as the statistical uncertainties. Although the central value of $Y_P$ is low, it is only slightly more than $1\sigma$ below the SBBN-predicted central value (see **Figure 10**). Because systematic errors dominate, it is unlikely that the errors are Gaussian distributed.

Alternatively, the recent analyses (e.g., OS, FK, PLP07), although differing among themselves in detail, are in agreement on the weighted mean of the post-BBN abundance, which can be used to provide an upper bound to $Y_P$. To this end, the PLP07 result is adopted as an alternate constraint on (an upper bound to) $Y_P$: $Y_P < 0.251 \pm 0.002$.

## 3.4. Lithium-7

Outside of the Sun, the solar system, and the local ISM (which are all chemically evolved), lithium has been observed only in the absorption spectra of very old, very metal-poor stars in the halo of the Galaxy or in similarly metal-poor galactic globular cluster (GGC) stars. These metal-poor targets are, of course, ideal for probing the primordial abundance of lithium. Even though lithium is easily destroyed in the hot interiors of stars, theoretical expectations supported by observational data suggest that although lithium may have been depleted in many stars, the overall trend is that its galactic abundance has increased with time (see Section 3). Therefore, to probe the BBN yield of $^7$Li, the key data are from the oldest, most metal-poor halo or GGC stars in the Galaxy (expected to form a plateau in a plot of Li/H versus Fe/H) such as those shown at low metallicity in **Figure 12**.

As for $^4$He, the history of the relic $^7$Li abundance determinations is an interesting and, perhaps, cautionary tale. For example, using a set of the lowest-metallicity halo stars, Ryan et al. (30) claimed evidence for a 0.3 dex increase in the lithium abundance, $[\mathrm{Li}] \equiv 12 + \log(\mathrm{Li/H})$, as the metallicity (measured logarithmically by the iron abundance relative to solar) increased over the range $-3.5 \leq [\mathrm{Fe/H}] \leq -1$. From this trend, they derived a primordial abundance of $[\mathrm{Li}]_P \approx 2.0-2.1$. This abundance is low compared to an earlier estimate of Thorburn's (31), who found $[\mathrm{Li}]_P \approx 2.25 \pm 0.10$. One source of systematic errors is the stellar temperature scale, which plays a key role in the connection between the observed equivalent widths and the inferred $^7$Li abundance.

Studies of halo and GGC stars employing the infrared flux method effective temperature scale suggested a higher lithium plateau abundance (32), $[\mathrm{Li}]_P = 2.24 \pm 0.01$, similar to Thorburn's (31) value. Melendez & Ramirez (33) reanalyzed 62 halo dwarfs using an improved infrared flux method effective temperature scale, failing to confirm the [Li]-[Fe/H] correlation claimed by Ryan et al. (30) and finding a higher relic lithium abundance, $[\mathrm{Li}]_P = 2.37 \pm 0.05$. In a very detailed and careful reanalysis of extant observations, with great attention to systematic uncertainties and the error budget, Charbonnel & Primas (34) also found no convincing evidence for a lithium trend with metallicity, and derived $[\mathrm{Li}]_P = 2.21 \pm 0.09$ for their full sample and $[\mathrm{Li}]_P = 2.18 \pm 0.07$ when they restricted their sample to unevolved (dwarf) stars. They suggested the Melendez-Ramirez value should be corrected downward

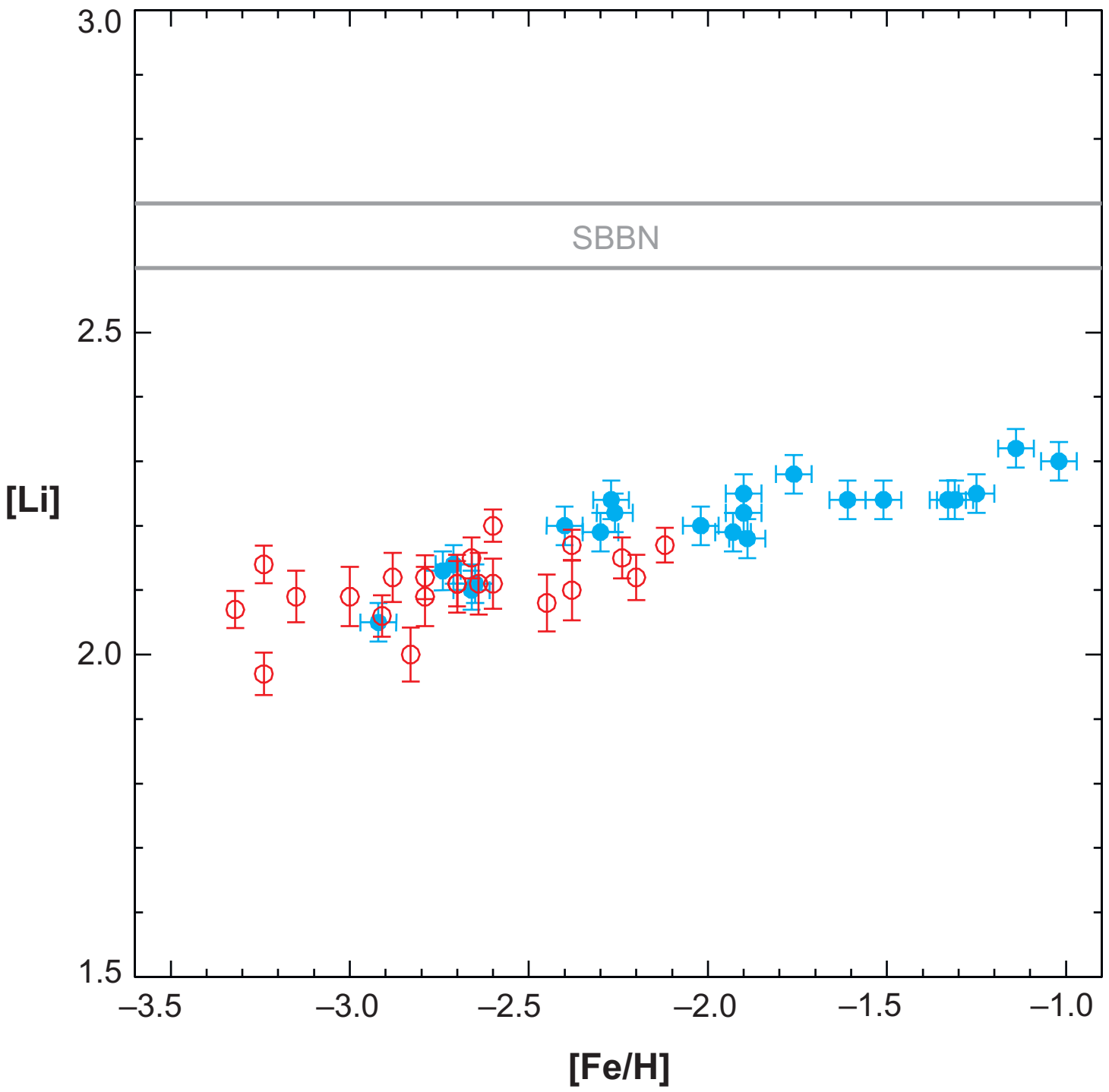

**Figure 12**

Lithium abundances, [Li] ≡ 12+ log(Li/H), versus metallicity (on a log scale relative to solar) from Reference 30 (*open red circles*) and Reference 35 (*filled blue circles*). The SBBN-predicted 1$\sigma$ band (see Section 2.2) is shown for comparison.

by 0.08 dex to account for different stellar atmosphere models, thereby bringing it into closer agreement with their results.

More recently, Asplund et al. (35) obtained and analyzed data for 24 metal-poor halo stars (filled circles in **Figure 12**), taking special care to address many stellar and atomic physics issues related to the path from the data to the abundances. The great surprise of their results is the apparent absence of a lithium plateau; at least from their data, the lithium abundance appears to continue to decrease with decreasing metallicity. Inclusion of Ryan et al.'s data at low metallicity flattens this trend, suggesting that $[\mathrm{Li}]_{\mathrm{P}} \approx 2.1 \pm 0.1$. There is clearly tension (if not outright conflict) between this estimate and the SBBN-predicted relic abundance of $[\mathrm{Li}]_{\mathrm{P}} = 2.65^{+0.05}_{-0.06}$ (see Section 2.2). Asplund et al. identify and discuss in some detail several possible sources of systematic errors: systematic errors in the abundance analysis; dilution and destruction of $^7$Li in the very old, very metal-poor stars; and uncertain or erroneous nuclear reaction rates (resulting in an incorrect SBBN-predicted abundance). Asplund et al. argue that it is unlikely that the abundance analysis errors can be large enough to bridge the gap, and Cyburt et al. (36) have used the observed solar neutrino flux to limit the nuclear reaction rate uncertainty, removing that too as a likely explanation. However, because the low-metallicity halo or GGC stars used to constrain primordial lithium are among the oldest in the Galaxy, they have had the most time to alter their surface lithium abundances by dilution and/or destruction, as is known to be important for many younger, higher-metallicity stars such as the Sun. Although mixing stellar surface material to the interior (or interior material to the surface) would destroy or dilute any prestellar lithium, the small dispersion observed among the low-metallicity halo star lithium abundances (in contrast to the very large spread for the higher-metallicity stars) suggests this correction may not be large enough ($\lesssim$0.1–0.2 dex at most) to bridge the gap between theory and observation; see, for example, Pinsonneault et al. (37) and further references therein.

In this context, Korn et al.'s (38) recent observations, coupled with stellar modeling, are of special interest. Korn et al. have observed stars in the globular cluster NGC 6397. The advantage of this approach is that these stars are (or should be) the same age and metallicity. They compare their observations of lithium and iron to models of stellar diffusion, finding evidence that both lithium and iron have settled out of the atmospheres of these old stars. Applying their stellar models to the data they infer for the unevolved abundances, [Fe/H] = –2.1 and [Li] = 2.54 ± 0.10, in excellent agreement with the SBBN prediction. More such data are eagerly anticipated.

### 3.5. Adopted Primordial Abundances

The discussion above reveals that large uncertainties in the relic abundances, inferred from the observational data, persist in this era of precision cosmology. Much work remains to be done by observers and theorists alike. At present, the relic abundance of D (the baryometer of choice) appears to be quite well constrained. Because observations of $^3$He are limited to the chemically evolved Galaxy, uncertain corrections to the zero-metallicity abundance are the largest source of uncertainty in its primordial abundance. Although there are a wealth of data on the abundances of $^4$He and $^7$Li in metal-poor astrophysical sites, systematic corrections are the sources of the largest uncertainties for these nuclides. For this reason, an alternate abundance is considered for each of them (shown in parentheses below). With this in mind, the following primordial abundances are adopted for the comparison between the observations and the theoretical predictions:

$$10^5(\mathrm{D/H})_\mathrm{P} \equiv y_\mathrm{DP} = 2.68^{+0.27}_{-0.25}, \quad 22.$$

$$10^5(^3\mathrm{He/H})_\mathrm{P} \equiv y_\mathrm{3P} = 1.1 \pm 0.2, \quad 23.$$

$$\mathrm{Y_P} = 0.240 \pm 0.006\ (<0.251 \pm 0.002), \quad 24.$$

$$12 + \log(\mathrm{Li/H})_\mathrm{P} \equiv 2 + \log(y_\mathrm{Li})_\mathrm{P} \equiv [\mathrm{Li}]_\mathrm{P} = 2.1 \pm 0.1 (2.5 \pm 0.1). \quad 25.$$

## 4. CONFRONTATION OF THEORY WITH DATA

In the context of the standard models of cosmology and particle physics, the BBN-predicted abundances of the light nuclides depend on only one free parameter, the baryon abundance $\eta_{10}$. Consistency of SBBN requires there be a unique value (or a limited range) of $\eta_{10}$ for which the predicted and observationally inferred abundances of D, $^3$He, $^4$He, and $^7$Li agree. If they agree, then in this era of precision cosmology it is interesting to ask if this value/range of $\eta_{10}$ is in agreement with that inferred from non-BBN-related observations of the CMB and the growth of LSS. If they do not agree, the options may be limited only by the creativity of cosmologists and physicists. It is, of course, interesting to ask if any challenges to the standard model(s) may be resolved through a nonstandard expansion rate ($S \neq 1$) or a lepton asymmetry ($\xi_e \neq 0$).

## 4.1. Standard Big Bang Nucleosynthesis

Let us first put SBBN to the test. In **Figure 13**, the SBBN-predicted values of $\eta_{10}$ (and the $1\sigma$ ranges) corresponding to the observationally inferred abundances adopted in Section 3.5 are shown. Although the SBBN-predicted lithium abundance is a double-valued function of $\eta_{10}$ (see **Figure 5**), only the higher-$\eta_{10}$ branch is plotted here. From D, the baryometer of choice, we find $\eta_D = 6.0 \pm 0.4$. Although this value, accurate to ~6%, is in excellent agreement with that inferred from the less well-constrained abundance of $^3$He ($\eta_3 = 5.6^{+2.2}_{-1.4}$), the abundances of $^4$He and $^7$Li correspond to very different—much smaller—values of the baryon abundance ($\eta_{He} = 2.7^{+1.2}_{-0.9}$, $\eta_{Li} = 4.0 \pm 0.6$). Because the corresponding values of $\eta_{10}$ are outside the ranges of applicability of the fits described in Section 2.2, the values of $\eta_{He}$ and $\eta_{Li}$ in **Figure 12** are from a numerical BBN code. However, for the central value of the D-predicted baryon abundance, the SBBN-predicted helium abundance is $Y_P = 0.248$, only $1.3\sigma$ away from the observationally inferred value. The lithium abundance poses a greater challenge; for $\eta_{10} \equiv \eta_D = 6.0$, the SBBN-predicted lithium abundance, $[\text{Li}]_P = 2.6$, is far from the observationally inferred value of $[\text{Li}]_P = 2.1 \pm 0.1$.

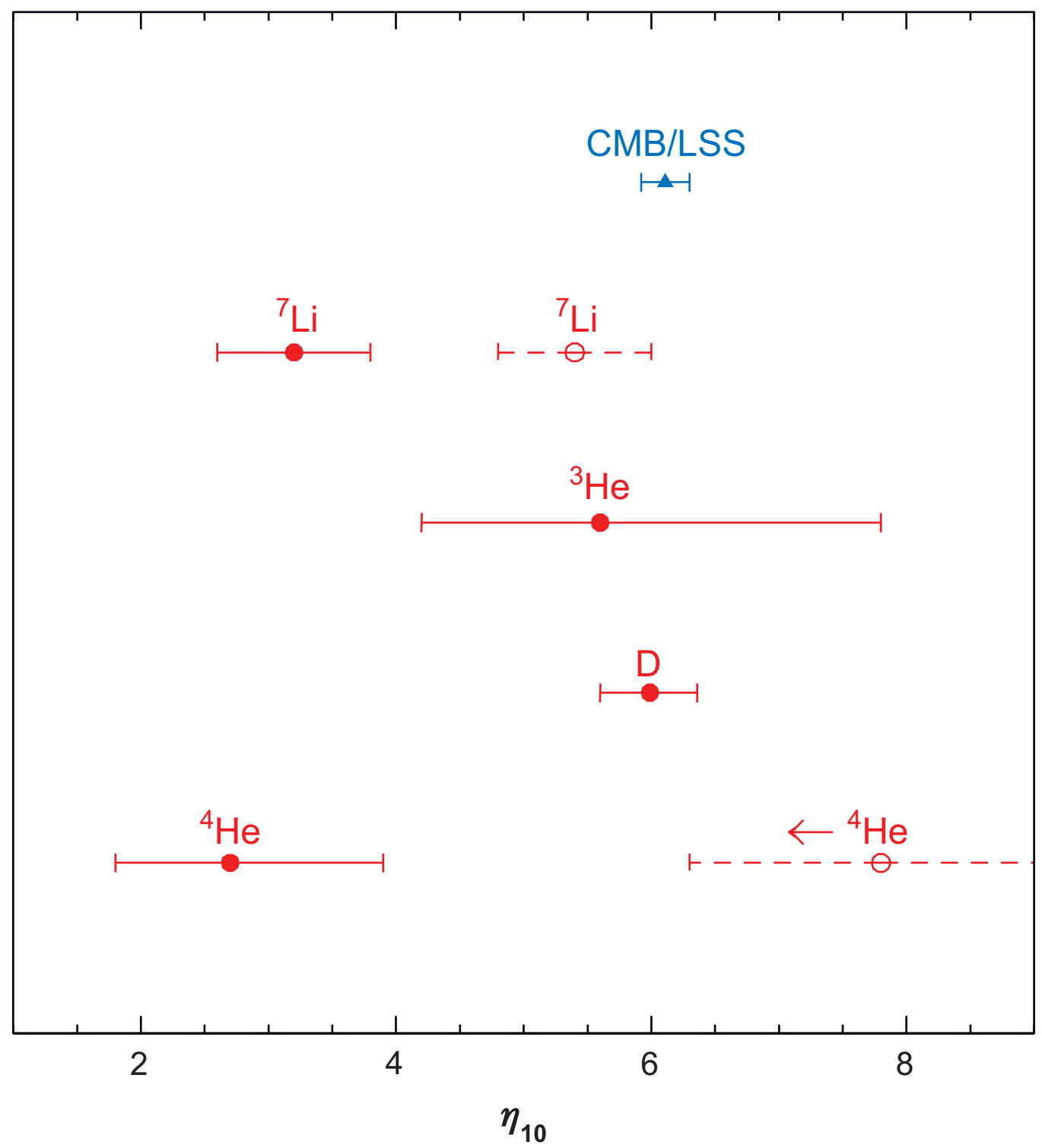

**Figure 13**
The SBBN-predicted values of $\eta_{10}$, and their $1\sigma$ uncertainties (*red filled circles*), corresponding to the primordial abundances adopted in Section 3.5, and the non-BBN value inferred from cosmic microwave background radiation (CMB) and large scale structure (LSS) data (*blue triangle*). The open circles and dashed lines correspond to the alternate abundances proposed for $^4$He and $^7$Li in Section 3.5.

It is interesting that the two nuclides that may pose challenges to SBBN are those for which systematic corrections may change the inferred relic abundances by the largest amounts. The alternate choices considered in Section 3.5, shown by the open circles in **Figure 13**, are in much better agreement with the D and $^3$He determined baryon abundance ($\eta_{\text{He}} < 7.8^{+1.9}_{-1.5}$, $\eta_{\text{Li}} = 5.4 \pm 0.6$).

Observations of the small temperature fluctuations in the CMB and of the LSS they seeded currently provide the tightest constraints on the universal abundance of baryons: $\eta_{\text{CMB/LSS}} = 6.11 \pm 0.20$ (2, 3), as shown in **Figure 13**. The observationally inferred relic abundances of D and $^3$He are in excellent agreement with the SBBN predictions for this value/range of $\eta_{10}$. Depending on the outcome of various systematic corrections for the primordial abundances of $^4$He and $^7$Li, the observationally inferred abundances may or may not pose a challenge to SBBN.

## 4.2. Nonstandard Big Bang Nucleosynthesis

Whether or not the relic abundances of $^4$He and/or $^7$Li are consistent with SBBN, the relic abundances, especially of D and $^4$He, can provide bounds on the parameters of some general classes of nonstandard physics. The data may be used in two ways to reveal such constraints. For example, ignoring the non-BBN (CMB/LSS) estimate for $\eta_{10}$, the primordial abundances of D and $^4$He may be used to constrain $S$ ($N_\nu$) or $\xi_e$:

$$\eta_{\text{He}} - \eta_D = -106(S - 1) = -\frac{579}{4}\xi_e. \qquad 26.$$

For the abundances choices (alternate) in Section 3.5,

$$S = 0.952 \pm 0.036 \; (<1.017 \pm 0.013), \; \xi_e = 0.034 \pm 0.026 \; (>-0.012 \pm 0.009). \qquad 27.$$

The constraint on $S$ corresponds to $N_\nu = 2.42^{+0.43}_{-0.41}$ ($N_\nu < 3.21 \pm 0.16$ for the alternate $^4$He abundance), whereas the $2\sigma$ range is $1.61 \leq N_\nu \leq 3.30$ ($N_\nu < 3.54$). In either case, the central values are less than $\sim 1.5\sigma$ away from SBBN. Indeed, it is interesting that the $2\sigma$ upper bound to the effective number of neutrinos at BBN is $N_\nu \leq 3.3$ ($N_\nu < 3.5$), constraining the presence at BBN of even one additional active or fully mixed neutrino. $N_\nu \leq 3.3 (<3.5)$ also precludes the existence at that time of a thermalized, light (relativistic) scalar for which the equivalent number of neutrinos corresponds to $\Delta N_\nu = 4/7$. For $N_\nu = 3$, the constraint on $S$ bounds the ratio of the early Universe gravitational constant to its present value $G'_N/G_N = 0.91 \pm 0.07$ or, at $2\sigma$, $0.77 \leq G'_N/G_N \leq 1.05$ ($< 1.09$). Note that although allowing $S \neq 1$ and/or $\xi_e \neq 0$ can reconcile the BBN-predicted and observed relic abundances of D and $^4$He, even if all three ($\eta_{10}$, $S$, $\xi_e$) are allowed to vary, there is no combination that can resolve the lithium problem. As seen from Equations 14 and 18, the non-SBBN-predicted abundance of lithium remains very close to its SBBN-predicted value.

Very similar, nearly identical, constraints are found if in place of the D abundance to constrain $\eta_{10}$, the CMB/LSS value of $\eta_{10}$ is employed. Here, too, the non-SBBN value of the lithium abundance remains very close to its SBBN-predicted value.

## 5. SUMMARY

Some 20 years ago, an article, "Big Bang Nucleosynthesis: Theories and Observations," by A.M. Boesgaard and the current author appeared in the 1985 issue of the *Annual Review of Astronomy and Astrophysics* (39). It is interesting to compare the conclusions from that review to those reached in 2006. At that time, the only data on D came from the solar system and the local ISM of the Galaxy. As a result, potentially large and uncertain evolutionary corrections were a barrier to constraining its primordial abundance, leading to the very large range in the relic abundance adopted there: $1-2 < y_{DP} < 20$. With such a large uncertainty in the baryometer of choice, all that could be inferred about the baryon abundance was that $\eta_{10} < 7-10$, consistent with our present estimate. Although there were no observations of $^3$He outside of the solar system at that time, Boesgaard & Steigman (39) noted, following Yang et al. (40), that solar system observations of D and $^3$He, in addition to some very general assumptions about the post-BBN evolution of D and $^3$He, suggested a bound on the sum of the relic abundances of D and $^3$He: $y_{DP} + y_{3P} < 6-10$, leading to a lower bound on the baryon abundance of $\eta_{10} > 3-4$, again consistent with our present estimate. As for $^4$He, we argued that all high-quality data were consistent with $Y_P = 0.24 \pm 0.02$, and we cautioned that "systematic effects—and not statistical uncertainties—dominate." *Plus ça change* .... Then, as now, the new (at that time) data on lithium in metal-poor stars favored $y_{Li} = 0.7-1.8$, corresponding to a very low SBBN baryon abundance of $1.6 \lesssim \eta_{10} \lesssim 4.0$. Twenty years ago, the neutron lifetime was quite uncertain and the central value and range [$900 \lesssim \tau_n(1985) \lesssim 935$] were significantly different from the currently favored value [$\tau_n = 885.7 \pm 0.8$ (8)]. Accounting for this difference, the Boesgaard-Steigman-quoted bound (39) on the effective number of neutrinos, $N_\nu < 3.8$ (for an assumed upper bound of $Y_P < 0.254$ and a lower bound on the baryon abundance of $\eta_{10} > 3$), translates to a present-day limit of $N_\nu < 4.0$.

The past 20 years have seen dramatic increases in the amount and the precision of cosmological data. Non-BBN data from the CMB and LSS (2, 3) constrain the baryon abundance to high accuracy ($\eta_{10} = 6.11 \pm 0.20$), leading to very accurate SBBN predictions of the relic abundances of D, $^3$He, $^4$He, and $^7$Li (see Section 2.2). A wealth of observational data has extended our reach on the primordial abundances of these light nuclides. These high-precision data reinforce the importance of accounting for systematic uncertainties if they are to be transformed into accurate relic abundance estimates. At present, the SBBN-predicted and observationally inferred primordial abundances of D and $^3$He are in excellent agreement, providing support for the standard models of cosmology and particle physics. The standard cosmological model provides a concordant description of the Universe at a few minutes and 400,000 years. Although it has profited from more high-precision data, the $^4$He abundance continues to be dominated by systematic uncertainties. Nonetheless, the predicted and observed $^4$He abundances agree within $2\sigma$, leading to strong constraints on new physics [$1.6 \lesssim N_\nu \lesssim 3.3$ ($N_\nu \leq 3.5$); $-0.027 \lesssim \xi_e \lesssim 0.086$ ($\xi_e \geq -0.030$); see Section 4.2]. BBN continues to provide a unique window on the early evolution of the Universe and on its particle content.

## DISCLOSURE STATEMENT

The author is not aware of any biases that might be perceived as affecting the objectivity of this review.

## ACKNOWLEDGMENTS

I am appreciative of the insights and advice provided over the years by my many collaborators on the topic of this review. This review profited from helpful discussions with Manuel Peimbert, Marc Pinsonneault, Donatella Romano, and Monica Tosi. I wish to thank Mandeep Gill for a careful reading of this manuscript. I owe a debt of gratitude to the copyeditor and the referee for useful suggestions that improved the accuracy and clarity of this manuscript. This research is supported at The Ohio State University by a grant from the U.S. Department of Energy (DE-FG02-91ER40690).

## LITERATURE CITED

1. Lesgourgues J, Pastor S. *Phys. Rep.* 429:307 (2006); Hannestad S. *Annu. Rev. Nucl. Part. Sci.* 56:137 (2006)
2. Spergel DN, et al. *Astrophys. J. Suppl.* 170:377 (2007)
3. Tegmark M, Eisenstein DJ, Strauss MA, Weinberg DH. *Phys. Rev. D* 74:123507 (2006)
4. Steigman G. *Nature* 261:479 (1976)
5. Yang J, Schramm DN, Steigman G, Rood RT. *Astrophys. J.* 227:696 (1979)
6. Kang H-S, Steigman G. *Nucl. Phys. B* 372:494 (1992)
7. Kernan P, Krauss L. *Phys. Rev. Lett.* 72:3309 (1994); Copi C, Schramm DN, Turner MS. *Science* 267:192 (1995); Coc A, et al. *Astrophys. J.* 600:544 (2004); Cyburt RH. *Phys. Rev. D* 70:023505 (2004); Serpico PD, et al. *JCAP* 0412:010 (2004)
8. Yao W-M, et al. *J. Phys. G Nucl. Part. Phys.* 33:1 (2006)
9. Mangano G, et al. *Nucl. Phys. B* 729:221 (2005)
10. Kneller JP, Steigman G. *New J. Phys.* 6:117 (2004)
11. Epstein RJ, Lattimer J, Schramm DN. *Nature* 263:198 (1976)
12. Rood RT, Steigman G, Tinsley BM. *Astrophys. J. Lett.* 207:L57 (1976); Olive KA, et al. *Astrophys. J.* 444:680 (1995); Dearborn DSP, Steigman G, Tosi M. *Astrophys. J.* 465:887 (1996). Erratum. *Astrophys. J.* 473:570 (1996); Galli D, Stanghellini L, Tosi M, Palla F. *Astrophys. J.* 477:218 (1997); Olive KA, Schramm DN, Scully ST, Truran J. *Astrophys. J.* 479:752 (1997); Palla F, et al. *Astron. Astrophys.* 355:69 (2000); Chiappini C, Renda A, Matteucci F. *Astron. Astrophys.* 395:789 (2002)
13. Steigman G, Walker TP. *Astrophys. J. Lett.* 385:L13 (1992)
14. Geiss J, Gloeckler JG. *Space Sci. Rev.* 84:239 (1998)
15. Linsky J, et al. *Astrophys. J.* 647:1106 (2006)
16. Steigman G. *MNRAS* 269:L53 (1994)
17. Kirkman D, et al. *Astrophys. J. Suppl.* 149:1 (2003)
18. O'Meara JM, Burles S, Prochaska JX, Prochter GE. *Astrophys. J.* 649:L61 (2006)
19. Bania TM, Rood RT, Balser DS. *Nature* 415:54 (2002)

20. Romano D, Tosi M, Matteucci F, Chiappini C. *MNRAS* 346:295 (2003)
21. Izotov YI, Thuan TX. *Astrophys. J.* 500:188 (1998); Izotov YI, Thuan TX. *Astrophys. J.* 602:200 (2004)
22. Davidson K, Kinman TD. *Astrophys. J.* 58:321 (1985)
23. Olive KA, Skillman ED. *New Astron.* 6:119 (2001); Olive KA, Skillman ED. *Astrophys. J.* 617:29 (2004)
24. Peimbert M, Peimbert A, Ruiz MT. *Astrophys. J.* 541:688 (2000); Peimbert A, Peimbert M, Luridiana V. *Astrophys. J.* 565:668 (2002); Luridiana V, Peimbert A, Peimbert M, Cerviño M. *Astrophys. J.* 592:846 (2003)
25. Steigman G, Viegas SM, Gruenwald R. *Astrophys. J.* 490:187 (1997)
26. Viegas SM, Gruenwald R, Steigman G. *Astrophys. J.* 531:813 (2000); Gruenwald R, Steigman G, Viegas SM. *Astrophys. J.* 567:931 (2002); Sauer D, Jedamzik K. *Astron. Astrophys.* 381:361 (2002)
27. Fukugita M, Kawasaki M. *Astrophys. J.* 646:691 (2006)
28. Peimbert M, Luridiana V, Peimbert A, Carigi L. astro-ph/0701313 (2007); Peimbert M, Luridiana V, Peimbert A. astro-ph/0701580 (2007)
29. Pagel BEJ, Simonson EA, Terlevich RJ, Edmunds MJ. *MNRAS* 255:325 (1992); Olive KA, Skillman ED, Steigman G. *Astrophys. J.* 489:1006 (1997); Izotov YI, Thuan TX. *Astrophys. J.* 500:188 (1998); Olive KA, Steigman G, Walker TP. *Phys. Rep.* 333:389 (2000); Luridiana V, Peimbert A, Peimbert M, Cerviño M. *Astrophys. J.* 592:846 (2003); Izotov YI, Thuan TX. *Astrophys. J.* 602:200 (2004); Olive KA, Skillman ED. *Astrophys. J.* 617:29 (2004); Fukugita M, Kawasaki M. *Astrophys. J.* 646:691 (2006)
30. Ryan SG, Norris JE, Beers TC. *Astrophys. J.* 523:654 (1999); Ryan SG, et al. *Astrophys. J.* 530:L57 (2000)
31. Thorburn JA. *Astrophys. J.* 421:318 (1994)
32. Bonifacio P, Molaro P. *MNRAS* 285:847 (1997); Bonifacio P, Molaro P, Pasquini L. *MNRAS* 292:L1 (1997)
33. Melendez J, Ramirez I. *Astrophys. J.* 615:L33 (2004)
34. Charbonnel C, Primas F. *Astron. Astrophys.* 442:961 (2005)
35. Asplund M, et al. *Astrophys. J.* 644:229 (2006)
36. Cyburt RH, Fields BD, Olive KA. *Phys. Rev. D* 69:123519 (2004)
37. Pinsonneault MH, Steigman G, Walker TP, Narayanan VK. *Astrophys. J.* 574:398 (2002)
38. Korn AJ, et al. *Nature* 442:657 (2006)
39. Boesgaard AM, Steigman G. *Annu. Rev. Astron. Astrophys.* 23:319 (1985)
40. Yang J, et al. *Astrophys. J.* 276:92